\documentclass[twocolumn]{aastex61}
\pdfoutput=1
\usepackage{amsmath,amstext}
\usepackage[T1]{fontenc}
\usepackage[figure,figure*]{hypcap}
\usepackage{tikz}
\usetikzlibrary{shapes,arrows}

\newcommand{\logg} {\log \textsl{\textrm{g}}}

\newcommand{\Te} {T_{\rm eff}}

\newcommand{\msun} {$M_\odot$}

\newcommand\gta{\lower 0.5ex\hbox{$\buildrel > \over \sim\ $}} 
\newcommand\lta{\lower 0.5ex\hbox{$\buildrel < \over \sim\ $}} 
\newcommand{\nh} {N({\rm H})/N({\rm He})}

\newcommand{\ha} {$\rm{H}{\alpha}$}

\shorttitle{WD Parameters in the Gaia Era}
\shortauthors{Bergeron et al.}

\begin{document}

\title{ON THE MEASUREMENT OF FUNDAMENTAL PARAMETERS OF WHITE DWARFS IN THE GAIA ERA}

\author{P. Bergeron, P. Dufour, G. Fontaine, S. Coutu, S. Blouin, C. Genest-Beaulieu, A. B\'edard, and B. Rolland}
\affiliation{D\'epartement de Physique, Universit\'e de Montr\'eal, C.P.~6128, Succ.~Centre-Ville, Montr\'eal, Qu\'ebec H3C 3J7, Canada; bergeron@astro.umontreal.ca, dufourpa@astro.umontreal.ca, fontaine@astro.umontreal.ca, coutu@astro.umontreal.ca, sblouin@astro.umontreal.ca, genest@astro.umontreal.ca, bedard@astro.umontreal.ca, rolland@astro.umontreal.ca}

\begin{abstract}

We present a critical review of the determination of fundamental
parameters of white dwarfs discovered by the {\it Gaia} mission. We
first reinterpret color-magnitude and color-color diagrams using
photometric and spectroscopic information contained in the Montreal
White Dwarf Database (MWDD), combined with synthetic magnitudes
calculated from a self-consistent set of model atmospheres with
various atmospheric compositions. The same models are then applied
to measure the fundamental parameters of white dwarfs using the
so-called photometric technique, which relies on the exquisite {\it
  Gaia} trigonometric parallax measurements, and photometric data from
Pan-STARRS, SDSS, and {\it Gaia}. In particular, we discuss at length
the systematic effects induced by these various photometric
systems. We then study in great detail the mass distribution as a
function of effective temperature for the white dwarfs
spectroscopically identified in the MWDD, as well as for the white
dwarf candidates discovered by {\it Gaia}. We pay particular attention
to the assumed atmospheric chemical composition of cool, non-DA
stars. We also briefly revisit the validity of the mass-radius relation
for white dwarfs, and the recent discovery of the signature of
crystallization in the {\it Gaia} color-magnitude diagram for DA white
dwarfs. We finally present evidence that the core composition of most of
these white dwarfs is, in bulk, a mixture of carbon and oxygen, an
expected result from stellar evolution theory, but never empirically
well established before. 

\end{abstract}


\section{INTRODUCTION} \label{intro}

The Sloan Digital Sky Survey (SDSS) has certainly been among the most
important developments in the last few years in terms of observational
data of white dwarf stars since the Palomar-Green survey
\citep{PG86}. The SDSS has been looking at 10,000 deg$^2$ of
high-latitude sky in five bandpasses ($ugriz$) and producing images in
these five bandpasses from which galaxies, quasars, and stars were
selected for follow-up spectroscopy.  The selection effects in this
survey are important, as discussed in \citet[][see also
  \citealt{EisensteinCat2006}]{Kleinman2004}, and therefore, it cannot
be considered as a complete survey in any sense. The number of white
dwarf stars discovered in the SDSS has rapidly grown from the first
Data Release \citep[2551 white dwarfs,][]{Kleinman2004} to well over
30,000 white dwarfs in Data Release 12 \citep{Kepler2016}.  Not only
has the SDSS significantly increased the number of spectroscopically
confirmed white dwarfs since the last published version of the McCook
\& Sion catalog \citep{mccook99}, but it has provided a phenomenal
source of homogeneous photometric observations in the $ugriz$ system
as well as optical spectroscopy for most objects. The SDSS has led to
numerous detailed spectroscopic analyses of white dwarfs, and in
particular for DA and DB stars (\citealt{EisensteinDB2006},
\citealt{Kepler2007}, \citealt{Tremblay2011}, \citealt{KK2015},
\citealt[][hereafter GBB19]{Genest2019}).

Another ongoing, and perhaps more important revolution in the white
dwarf field is the {\it Gaia} mission, which will probably discover
about 400,000 white dwarfs when the mission is completed, with a
detection probability of almost 100\% up to a distance of 100 pc
\citep{Jordan2007}. The {\it Gaia} Data Release 2 \citep[][hereafter
  GaiaHRD]{GaiaHRD} has already provided precise astrometric and
photometric data for $\sim$260,000 high-confidence white dwarf
candidates \citep{GentileFusillo2019}. These combined astrometric and
photometric data sets allow, for the first time, the measurements of
fundamental parameters (effective temperature, radii, and mass) of
large samples of white dwarf stars, using the so-called photometric
technique (\citealt{Hollands2018}, \citealt{Tremblay2019},
GBB19). Most white dwarf candidates identified by {\it Gaia} still
require spectroscopic observations and confirmations,
however. Furthermore, {\it Gaia} provides photometry only in wide
passbands --- $G$, $G_{\rm BP}$, and $G_{\rm RP}$, which are not the
most optimal set to be used for model atmosphere
modeling\footnote{Actually, the broadband filter $G$ is almost the sum
  of the $G_{\rm BP}$ and $G_{\rm RP}$ bandpasses, and in that sense,
  it does not represent an {\it independent} data point in the
  sampling of the spectral energy distribution.}. Fortunately, there
are now a growing number of all-sky (or almost) surveys with narrower
bands, such as Pan-STARRS (Panoramic Survey Telescope And Rapid
Response System; \citealt{Chambers2016}, \citealt{Tonry2012}).

Given that such large ensembles of combined astrometric (trigonometric
parallaxes in particular) and photometric data have become available
only recently, the fundamental parameters derived from the photometric
technique are not as mature and well understood as those obtained from
spectroscopy. To overcome this situation, we present in this paper a
comprehensive investigation of the physical parameters of white dwarfs
derived from the photometric technique. In Section \ref{sec:colmag},
we first investigate the classical color-magnitude and color-color
diagrams for white dwarfs, both observationally and theoretically. In
Section \ref{sec:param}, we discuss the determination of physical
parameters of white dwarfs using the photometric technique, with a
particular emphasis on the use of various photometric systems.
Finally, we explore in Section \ref{sec:implic} various implications
of our results for white dwarf physics, including a detailed study of
the mass distributions under various assumptions about the chemical
composition, and a closer look at the crystallization process. Our
discussion and conclusions follow in Section \ref{sec:disc}.

\section{COLOR-MAGNITUDE AND COLOR-COLOR DIAGRAMS}\label{sec:colmag}

\begin{figure}[t]
\centering
\includegraphics[width=\linewidth]{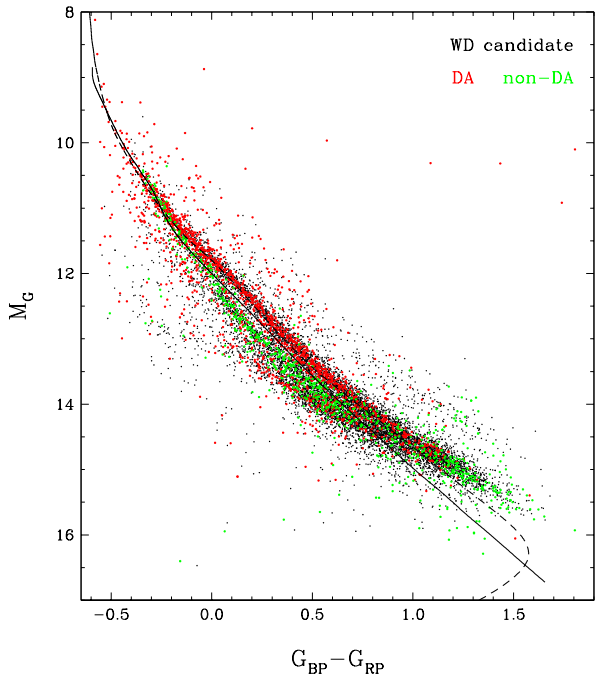}
\vspace*{-1cm}
\caption{$M_G$ vs $G_{\rm BP}-G_{\rm RP}$ color-magnitude diagram for
  white dwarfs in the MWDD with {\it Gaia} parallax measurements more
  precise than 10\%, and distances less than 100 pc. White dwarf
  candidates (8787 objects) are shown as small black dots, while
  spectroscopically identified DA (1827) and non-DA (843) stars are
  represented by larger red and green dots, respectively. Also shown
  are theoretical color sequences for 0.6 \msun\ white dwarfs with
  pure hydrogen (dashed line) and pure helium (solid line) atmospheric
  compositions.\label{color_mag}}
\end{figure}

The first Hertzsprung-Russell diagram for white dwarfs using data from
the {\it Gaia} mission has been presented in Figure 13 of GaiaHRD ---
in this case, $M_G$ vs $(G_{\rm BP}-G_{\rm RP})$. We show in Figure
\ref{color_mag} our own version of this color-magnitude diagram, but
only for objects with a distance $D< 100$ pc, based on {\it Gaia}
photometric and parallax data extracted from the Montreal White Dwarf
Database\footnote{http://montrealwhitedwarfdatabase.org/}
\citep{MWDD}. We use different color symbols to distinguish white
dwarf candidates from {\it Gaia}, as well as spectroscopically
identified DA and non-DA (DO, DB, DQ, DZ, DC, and all subtypes)
stars. Composite systems with a {\it known} M-dwarf companion are also
excluded from this diagram. For the selection of white dwarf
candidates, instead of using the extensive catalogue of
\citet{GentileFusillo2019}, we use the same cuts as in GaiaHRD (see
their Section 2.1), to which we add
$G-10+5\log_{10}\pi>10+2.6\ (G_{\rm BP}-G_{\rm RP})$ to select only
white dwarfs, a cut also suggested in GaiaHRD. We also restrict our
sample to $\sigma_\pi /\pi<0.1$.

Also reproduced in this color-magnitude diagram are theoretical color
sequences for 0.6 \msun\ white dwarfs with pure hydrogen and pure
helium atmospheric compositions, calculated using state-of-the-art
model atmospheres described in Section \ref{sec:param}. As discussed
in GaiaHRD, the most striking feature in this diagram is a clear
bifurcation between the DA and non-DA stars in the range $0.0<(G_{\rm
  BP}-G_{\rm RP})<0.8$. Although the pure hydrogen models follow
nicely the observed DA sequence, the pure helium models actually go
through the gap where the bifurcation between the DA and the DB stars
occurs. There are at least two interpretations that have been proposed
for these results, also discussed in GaiaHRD. The first obvious
interpretation of this bifurcation is atmospheric composition, which
differentiates the hydrogen-atmosphere DA white dwarfs from the
helium-atmosphere non-DA stars. In this case, it has been suggested
that perhaps the helium models simply fail to go through the observed
non-DA sequence at 0.6 \msun\ because of physical inaccuracies in the
model atmospheres. An alternative explanation that has been proposed
is related to differences in stellar mass, which affect the radius of
the star, and thus the absolute magnitude in such color-magnitude
diagrams. In this other case, the observed non-DA sequence in Figure
\ref{color_mag} could be interpreted as a higher-than-average mass for
these objects. We now explore these two interpretations further using
our full set of synthetic colors for white dwarfs.

\begin{figure}[t]
\centering
\includegraphics[clip=true,trim=2.5cm 4.5cm 2cm 3cm,width=\linewidth]{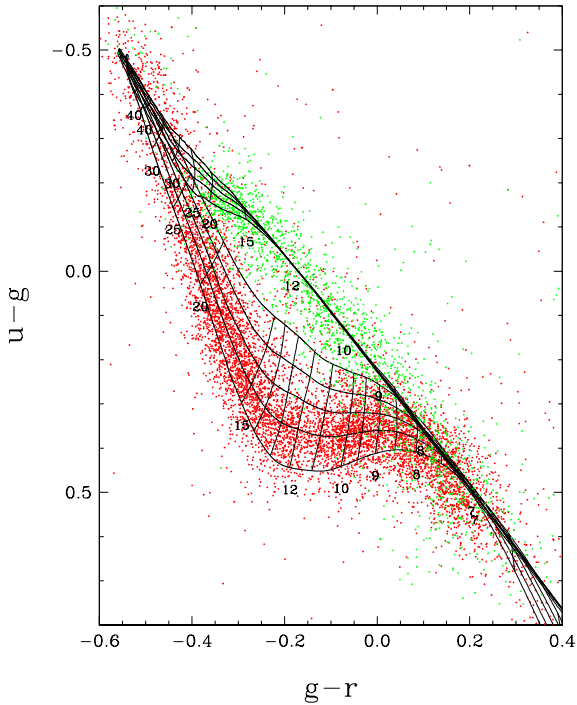}
\caption{SDSS $(u-g)$ vs $(g-r)$ two-color diagram for all
  spectroscopically confirmed white dwarfs in the MWDD, regardless of
  distance, with {\it Gaia} parallax measurements more precise than
  10\%. DA and non-DA (DO, DB, DQ, DZ, DC) white dwarfs (7935 DA and
  2463 non-DA stars) are represented by red and green dots,
  respectively. Theoretical colors for pure hydrogen (bottom) and pure
  helium (top) models are also shown with constant masses of 0.4, 0.6,
  0.8, 1.0, and 1.2 \msun, from bottom to top, and effective
  temperatures indicated in units of $10^3$~K.\label{color_color}}
\end{figure}

As can be seen in Figure \ref{color_mag}, pure hydrogen and pure
helium models overlap almost completely in color-magnitude diagrams
($\Delta M_G \lesssim 0.2$), except at low effective temperature where
collision-induced absorption (CIA) by molecular hydrogen becomes
important. However, the differences in colors between hydrogen- and
helium-atmosphere white dwarfs become much more important when the
Balmer jump is considered, which can be measured by observing the flux
on each side of the hydrogen photoionization threshold (from the $n=2$
level), located at $\lambda\sim 3640$ \AA. One of the best examples is
the color-color diagram built from SDSS magnitudes, $(u-g)$ vs
$(g-r)$, displayed in Figure \ref{color_color} (see also the $M_u$ vs
$(u-g)$ color-magnitude diagram shown in Figure 15 of
GaiaHRD). Because of the relatively small number of white dwarfs
within 100 pc with $ugriz$ data available, we include in this figure
objects at all distances (still with $\sigma_\pi /\pi<0.1$), and the
colors are therefore dereddened following the procedure described
in \citet{Harris2006} and discussed in Section
\ref{sec:phottech}. Also reproduced in Figure \ref{color_color} are
our theoretical colors for pure hydrogen and pure helium model
atmospheres at various temperatures and stellar masses. The important
effect produced by the Balmer jump, as measured by the $(u-g)$
color-index, is clearly visible between $\Te\sim8000$~K and 20,000~K
(see \citealt{Shipman1980} for a detailed physical explanation of the
$\logg$ dependence of this jump). More importantly, the theoretical
colors reproduce perfectly the observed sequences for both the DA and
non-DA white dwarfs in this diagram, and there is a sensitivity on the
mass for DA stars leading to inferred masses around 0.6
\msun, as expected. We conclude, at least based on this color-color
diagram, that there appears to be no major problems with the color
predictions from our model atmospheres.

\begin{figure*}[t]
\centering
\includegraphics[angle=270,clip=true,trim=2cm 1cm 3cm 0cm,width=\linewidth]{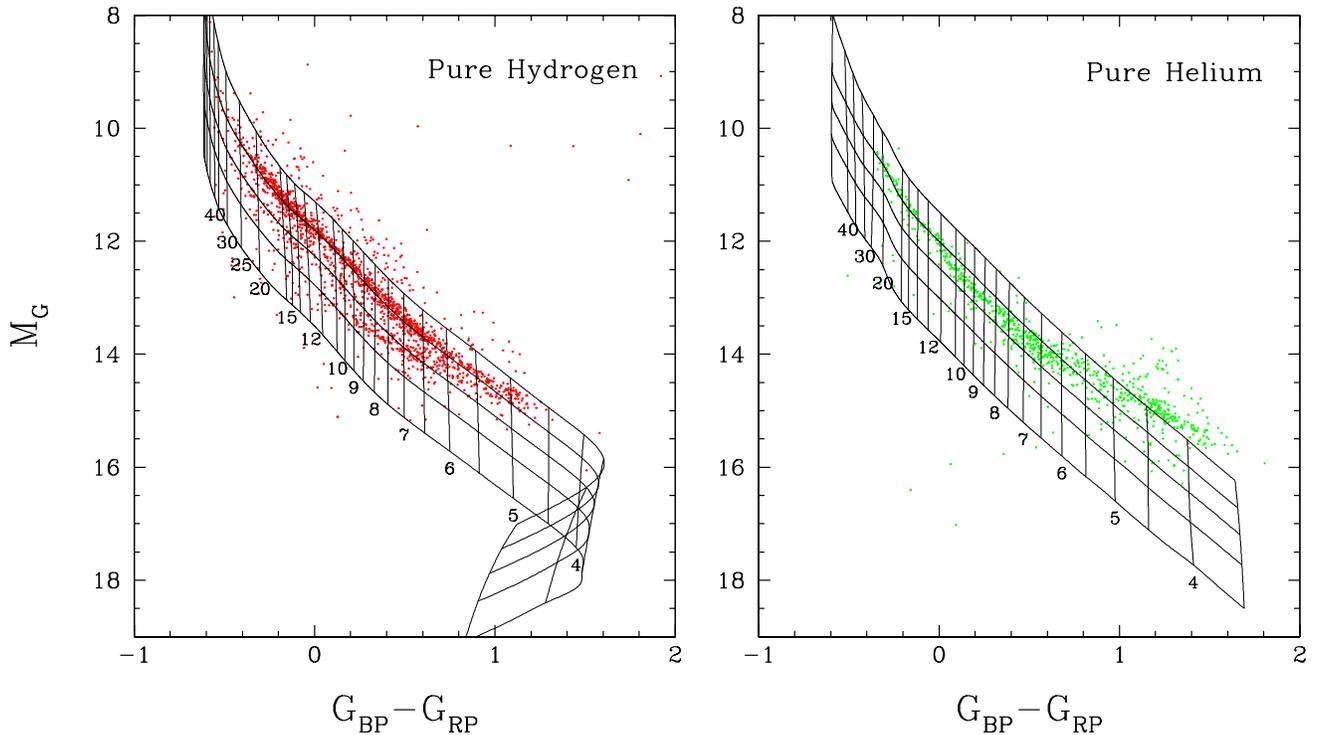}
\caption{$M_G$ vs $(G_{\rm BP}-G_{\rm RP})$ color-magnitude diagrams
  for DA (left; 1792 objects) and non-DA (right; 838 objects) white
  dwarfs in the MWDD with {\it Gaia} parallax measurements more
  precise than 10\% and distances less than 100 pc. Theoretical colors
  for pure hydrogen (left) and pure helium (right) models are also
  shown with constant masses of 0.4, 0.6, 0.8, 1.0, and 1.2 \msun,
  from top to bottom, and effective temperatures indicated in units of
  $10^3$~K.\label{color_mag_mass}}
\end{figure*}

In Figure \ref{color_mag_mass}, we show the same $M_G$ vs $(G_{\rm
  BP}-G_{\rm RP})$ color-magnitude diagram as before (with $D<100$
pc), but this time by splitting the DA and non-DA stars into two
panels, excluding the {\it Gaia} white dwarf candidates. We also
superpose in each panel the theoretical colors for the corresponding
pure hydrogen or pure helium atmospheric composition, and for various
values of effective temperature and stellar mass. For the DA stars
(left panel), the observed sequence follows the 0.6 \msun\ model
sequence almost perfectly (slightly lower than 0.6 \msun\ actually),
with a small departure towards lower masses below $\Te\sim5500$~K. We
can also identify a large population of extreme low-mass DA white
dwarfs ($M<0.4$ \msun) --- most likely unresolved double degenerate
binaries --- as well as a nearly horizontal sequence of massive DA
stars, which has recently been interpreted as evidence for
crystallization \citep{TremblayNat2019}. 

In the case of non-DA white dwarfs (right panel), the observed
sequence follows very nicely the 0.6 \msun\ model sequence from
$\Te\sim30,000$~K down to $12,000$~K, which corresponds to the region
where DB white dwarfs are located. We also find little evidence for
any large population of low- or high-mass white dwarfs in this
temperature range. Below 12,000~K, however, a significant departure
from the 0.6 \msun\ sequence can be observed, with a sudden and nearly
constant shift towards larger masses by an amount of $\sim$0.1 \msun,
down to $\Te\sim6000$~K.  Below $\sim$6000~K, the scatter in mass for
the non-DA stars increases considerably, with the average mass of the
bulk of objects decreasing steadily, even reaching values well below
0.4 \msun\ at $\Te\lesssim 4500$~K. We also find evidence again for
overluminous, unresolved double degenerates in this temperature
range. The most plausible explanation for these strange features is
that most cool non-DA white dwarfs in this diagram probably have
hydrogen-dominated atmospheres. Indeed, a closer inspection of the
results displayed in Figure \ref{color_mag} reveals that the coolest
non-DA white dwarfs represent a natural extension of the cool DA
sequence. It is thus possible that most of the coolest non-DA stars in
the right panel of Figure \ref{color_mag_mass} have hydrogen-dominated
atmospheres, and belong instead in the left panel with other
hydrogen-rich objects. We revisit this point further in Section
\ref{sec:gaiawd}.

Color-magnitude and color-color diagrams, despite their general
interest in terms of the global properties of the sample, contain
limited information in terms of interpreting the physical parameters
of white dwarfs, most importantly effective temperatures, stellar
masses, and atmospheric compositions. We now turn our attention to the
more physical approach of fitting spectral energy distributions with
the predictions of detailed model atmospheres.

\section{PHOTOMETRIC DETERMINATIONS OF PHYSICAL PARAMETERS}\label{sec:param}

\subsection{The Photometric Technique}\label{sec:phottech}

The photometric technique has been developed and first applied to
large ensembles of white dwarfs by \citet{BRL97}. Briefly, a set of
magnitudes are converted into average fluxes using appropriate zero
points, and fitted --- using a least-squares method --- with average
synthetic fluxes calculated from model atmospheres with the
appropriate chemical composition.  In this case, the fitted parameters
are the effective temperature, $\Te$, and the solid angle,
$\pi(R/D)^2$, where $R$ is the radius of the star, and $D$ its
distance from Earth. If the distance is known from the parallax
measurement, the radius can be measured directly, and converted into
stellar mass ($M$) using evolutionary models, which provide the
required temperature-dependent mass-radius relation. Further details
about this fitting method are given in GBB19, and will not be repeated
here. In particular, we use the same atmosphere and evolutionary
models as in GBB19, with the following exception. Here we rely on a
new grid of pure helium models computed using the atmosphere code
described in \citet{blouin2018model}. This code includes numerous
high-density effects relevant for the modeling of cool helium-rich
white dwarfs.  Among the most important effects considered, continuum
opacities are corrected for collective interactions
\citep{iglesias2002density,rohrmann2018rayleigh}, collision-induced
absorption from the He$-$He$-$He interaction is included
\citep{kowalski2014infrared}, an ab initio equation of state is used
\citep{becker2014ab}, and the ionization equilibrium of helium is
assessed from the ab initio calculations of
\citet{kowalski2007equation}.

GBB19 adopted the dereddening procedure outlined in \citet{Harris2006}
where the extinction is assumed to be negligible for stars with
distances less than 100 pc, to be maximum for those located at
$|z|>250$ pc from the galactic plane, and to vary linearly along the
line of sight between these two regimes. Although this approach is
probably valid for stars well above the galactic plane, as is the case
for the SDSS white dwarfs analyzed by GBB19, it is certainly not
appropriate for stars very close to the galactic plane ($z\sim 0$),
for which no reddening is predicted, even if the object lies at
extremely large distances from Earth. A more reasonable dereddening
procedure has been proposed by \citet[][see their Section
  4]{GentileFusillo2019}, which is applicable to all sky surveys, such
as the {\it Gaia} sample. Note that in the case of white dwarfs in the
SDSS, GBB19 demonstrated that both dereddening procedures yield
similar results (see their Figure 15). However, we found that in some
cases, the procedure proposed by Gentile Fusillo et al.~yields
spurious results when the star is nearby and that the maximum
extinction along the line of sight is large. One example is GD 50 (WD
0346$-$011) at a distance of only 31.2 pc, and a maximum extinction of
$E(B-V)=0.1601$ along the line of sight \citep{SF2011}. For this hot
DA star, we obtain a spectroscopic temperature of 42,670~K, and a
photometric temperature of 40,845~K based on PanSTARRS $grizy$
photometry, assuming the dereddening procedure of Harris et al.\ ---
no reddening in this case since $D<100$~pc. However, if we use the
procedure proposed by Gentile Fusillo et al., this photometric
temperature jumps to 59,090~K, i.e., more than 16,000~K above the
spectroscopic value.  We find similar problems with other nearby
objects in our sample.  As discussed by the authors, improved {\it
  Gaia} DR2 reddening maps in three dimensions will eventually
supersede these simple parameterizations, but in the meantime, we will
consider both dereddening procedures, depending on the white dwarf
sample analyzed.

A few additional details are also worth mentioning. Even though modern
magnitude measurements, such as SDSS or Pan-STARRS, are quoted with
extremely small uncertainties --- sometimes as small as millimagnitudes ---,
the conversion to average fluxes remains the largest source of
uncertainty when using the photometric technique \citep{HB2006}. For
example, while the Pan-STARRS photometric system attempts to be as
close as possible to the AB magnitude system, the Pan-STARRS
implementation has an accuracy of only $\sim$0.02 mag according to
\citet{Tonry2012}. Also, the SDSS magnitude system is not exactly on
the AB magnitude system either, and corrections must be added to the
$uiz$ bandpasses: $u_{\rm AB} = u_{\rm SDSS}-0.040$, $i_{\rm AB} =
i_{\rm SDSS}+0.015$, and $z_{\rm AB} = z_{\rm SDSS}+0.030$
\citep{EisensteinCat2006}. Vega-based magnitudes also suffer from similar
problems since more than often, the magnitude of Vega is not even
known on the given system. For these reasons, we always adopt in our
fitting procedure a {\it lower limit} of 0.03 mag uncertainty in all
bandpasses. This ensures that all bandpasses have more equal weights,
and that one magnitude with an extremely small error bar does not
drive the overall photometric solution.

\subsection{Adopted Photometric System}

One of the critical issues when using the photometric technique is to
identify which is the most optimal and reliable photometric system.
Obviously, the SDSS $ugriz$ photometry has proven to be very useful
and reliable (see GBB19 and references therein), but it is only
available for a portion of the sky, while magnitudes measured by {\it
  Gaia} and Pan-STARRS are almost all-sky surveys (three-quarters of
the entire sky for Pan-STARRS). \citet{GentileFusillo2019} have
investigated the use of several photometric systems, including {\it
  Gaia} ($G$, $G_{\rm BP}$, $G_{\rm RP})$, and found a generally good
agreement between the derived atmospheric parameters. In what follows,
we present our own assessment of the internal consistency between
effective temperature and mass values obtained from photometric fits
based on various photometric systems (PSF magnitudes). To do so, we
first rely on the spectroscopic sample of relatively bright DA stars
of \citet{Gianninas2011}. More specifically, we compare spectroscopic
temperatures and masses with those obtained from photometry, using a
procedure identical to that described in detail by GBB19 for DA stars
in the SDSS. We exclude from our sample all composite systems
containing a M-dwarf companion, as well as bright objects for which
the magnitudes are saturated (the SDSS photometry in particular).
Because, most of the DA stars in the Gianninas et al.~sample are
nearby objects, we adopt the dereddening procedure of
\citet{Harris2006} to avoid the problems discussed above. We point
out, however, that the dereddening procedure of
\citet{GentileFusillo2019} yields similar results, but many spurious
results are observed, such as the case of GD 50 mentioned in the
previous section.

\begin{figure*}[t]
\centering
\includegraphics[clip=true,trim=0cm 1.5cm 1cm 1cm,width=0.7\linewidth]{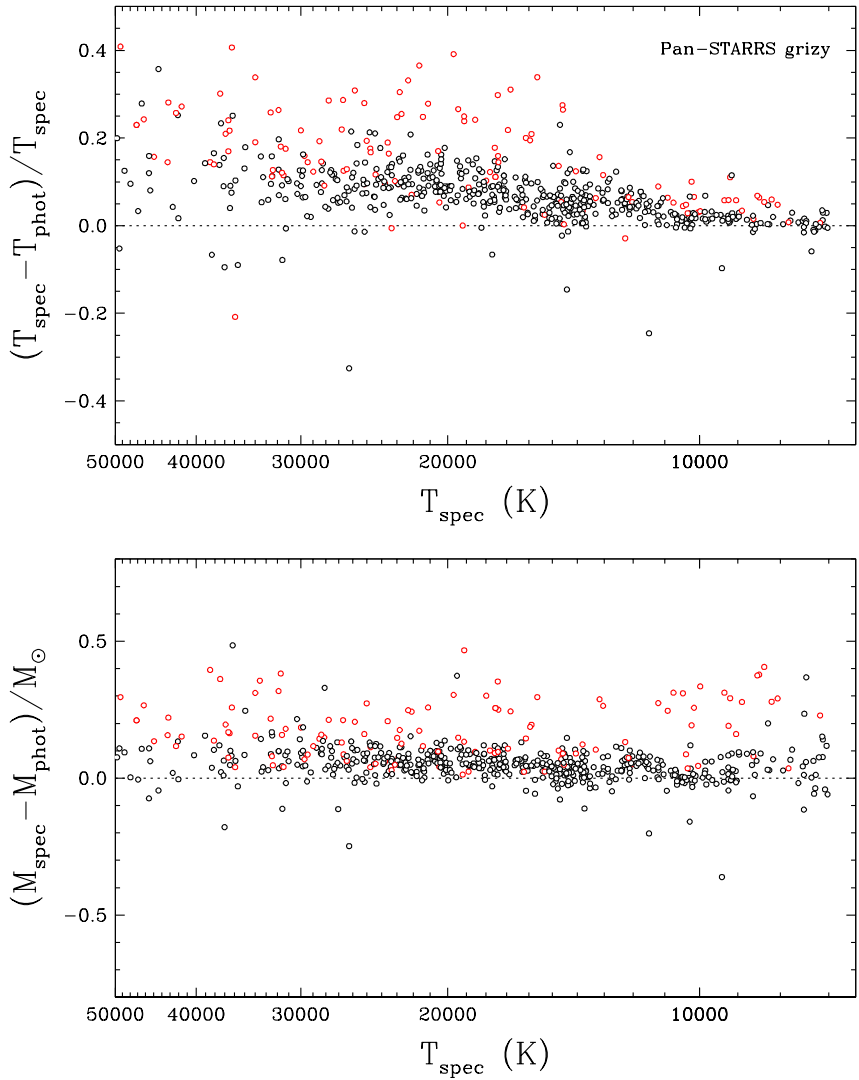}
\caption{Top panel: Differences between spectroscopic and photometric
  effective temperatures as a function of $\Te$ for DA stars drawn
  from the sample of \citet{Gianninas2011}, using photometric fits to
  the Pan-STARRS $grizy$ data. The dashed line indicates equal
  temperatures. The red symbols represent double degenerate binaries,
  or candidates, with photometric masses below 0.48 \msun. Bottom
  panel: Same results but for differences between spectroscopic and
  photometric masses.\label{comp_MASTERLIST_grizy}}
\end{figure*}

We show in the upper panel of Figure \ref{comp_MASTERLIST_grizy} the
differences between spectroscopic and photometric temperatures
measured from fits to Pan-STARRS $grizy$ photometry. The spectroscopic
solutions have been obtained using our grid of model atmospheres for
DA stars with the ML2/$\alpha=0.7$ version of the mixing-length
theory, and the 3D corrections\footnote{Note that these 3D corrections
  do not apply to the photometric solutions.} from \cite{Tremblay2013}
have been applied to both $T_{\rm eff}$ and $\log g$ values. Our
results can be contrasted with those obtained for DA stars in the SDSS
by GBB19 (see their Figures 13 and 17). While comparisons between
spectroscopic and photometric temperatures both show a systematic
offset --- with the spectroscopic temperatures being larger than the
photometric values ---, the differences observed here are much larger
(10 to 20\%) than those reported by GBB19 ($\lesssim$10\%). Note that
we exclude from this discussion the largest discrepant cases (up to
40\%), which occur for unresolved double degenerate binaries (or
binary candidates), shown by red symbols in Figure
\ref{comp_MASTERLIST_grizy}. More importantly, the comparison between
spectroscopic and photometric masses (bottom panel of Figure
\ref{comp_MASTERLIST_grizy}) shows a systematic offset of about 0.1
\msun, with spectroscopic masses being larger, while GBB19 (see their
Figure 17) find instead a much better agreement, with no apparent
systematic offset.

\begin{figure*}[t]
\centering
\includegraphics[clip=true,trim=0cm 1cm 1.5cm 0.5cm,width=0.75\linewidth]{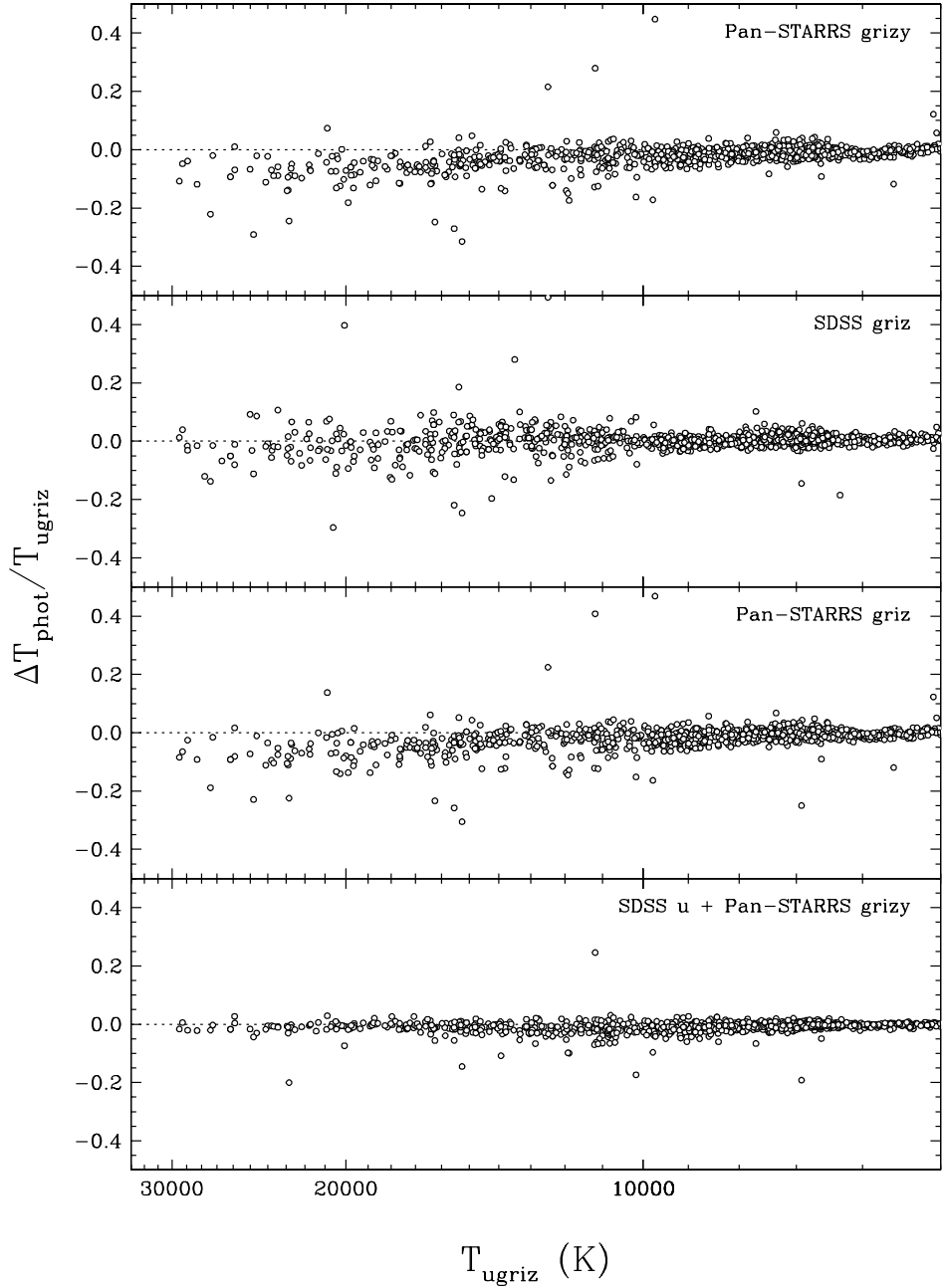}
\caption{Differences between photometric
  temperatures as a function of $T_{ugriz}$ --- the temperature
  obtained from SDSS $ugriz$ photometry --- for DA stars drawn from
  the SDSS with $D<100$ pc. $\Delta T_{\rm phot}$ represents the
  difference between the temperature obtained using the
  photometric data set indicated in each panel, and $T_{ugriz}$.  The dashed lines
  indicate equal temperatures. \label{comp_PHOT_all}}
\end{figure*}

Since the results of GBB19 are based on both a different white dwarf
sample (SDSS) and a different set of photometric measurements
($ugriz$), we explore here the same sample of DA stars drawn from the
SDSS, but we restrict our analysis to white dwarfs with $D<100$ pc to
minimize the effects of interstellar reddening. We also exclude all
objects with $M\lesssim0.48$ \msun, which are most likely unresolved
double degenerate binaries. Instead of using the spectroscopic
temperatures as a reference, we rely on photometric temperatures based
on SDSS $ugriz$ photometry ($T_{ugriz}$). The temperatures obtained
from various photometric systems and filter sets are compared with
$T_{ugriz}$ in Figure \ref{comp_PHOT_all}. The top panel shows the
comparison with Pan-STARRS $grizy$ photometry, also used in Figure
\ref{comp_MASTERLIST_grizy}.  Although the temperatures are in good
agreement below $\Te\sim15,000$~K, an increasing shift appears above
this temperature, with the SDSS $ugriz$ temperatures exceeding those
derived from Pan-STARRS $grizy$ ($\Delta T_{\rm
  phot}=T_{grizy}-T_{ugriz}$ in the upper panel of Figure
\ref{comp_PHOT_all}). Hence, had GBB19 used Pan-STARRS instead of SDSS
photometry, they would have found discrepancies between spectroscopic
and photometric temperatures that are much larger than those reported
in their Figures 13 and 17, and consistent with our results
displayed in Figure \ref{comp_MASTERLIST_grizy} using the Gianninas et
al.~sample.

Because the discrepancy between photometric temperatures becomes more
important at high temperatures, we first believed that the specific
use of the $u$ filter in the model fits based on SDSS $ugriz$
photometry was responsible for the observed shift in temperature. As discussed by
GBB19 (see in particular their Figure 4), the $u$ magnitude becomes
important when fitting hot white dwarfs that are in the Rayleigh-Jeans
regime. This is also illustrated in Figure 6 of \citet{Genest2014},
where photometric temperatures obtained using the full $ugriz$
photometric set are compared with those derived only from $griz$. To
test our hypothesis, we fitted the same sample of DA white dwarfs by
excluding the $u$ bandpass, thus fitting only the SDSS $griz$
photometry, the results of which are displayed in the second panel of
Figure \ref{comp_PHOT_all}. We can see that even though the scatter at
high temperatures has increased significantly, no systematic shift
appears, in sharp contrast with the results based on Pan-STARRS
photometry. \citet{Genest2014} reached a similar conclusion (see
their Section 2.4.2).

\begin{figure*}[t]
\centering
\includegraphics[angle=270,clip=true, trim=1cm 1.5cm 1cm 1cm, width=0.9\linewidth]{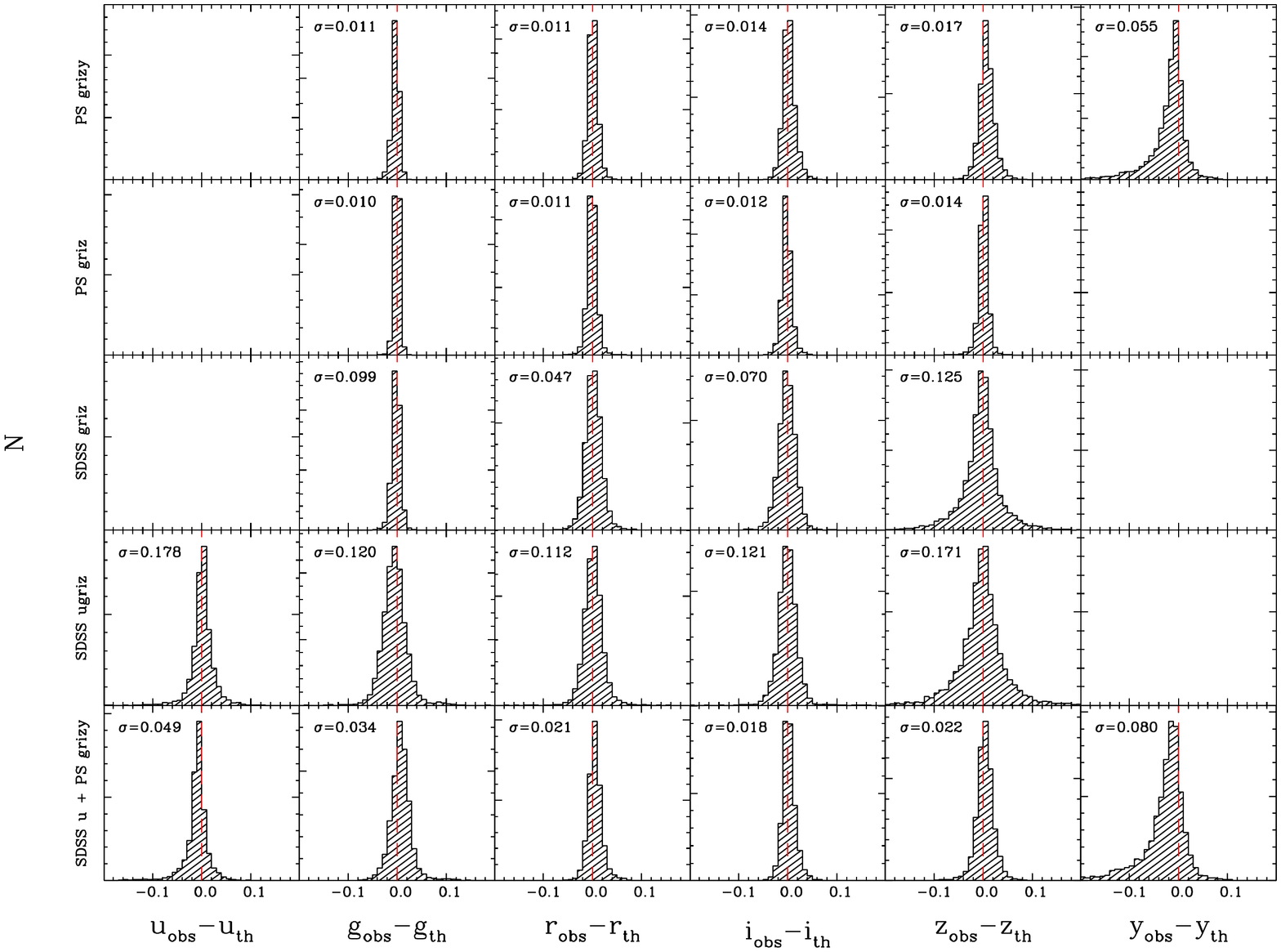}
\caption{Histograms showing the difference between the observed (obs)
  magnitudes and those predicted by the photometric technique (th) for
  various photometric systems and filters sets indicated on the left
  hand side of each row (PS stands for Pan-STARRS). The thick dashed
  lines correspond to $m_{\nu,{\rm obs}}=m_{\nu,{\rm th}}$. Also
  indicated in each panel is the 1$\sigma$ standard
  deviation.\label{diff_mag_all}}
\end{figure*}

To better understand these systematics differences, we performed the
same experiment as in Figure 8 of \citet{Genest2014}, which shows the
histogram distributions between observed (obs) and predicted
theoretical (th) magnitudes for each individual bandpass of the
$ugriz$ system. As discussed by the authors, if the photometric system
is well calibrated --- at least in a relative sense ---, all
histograms should appear symmetrical and centered on $m_{\nu,{\rm
    obs}}-m_{\nu,{\rm th}}=0$. We present in Figure \ref{diff_mag_all}
such histograms based on our own photometric fits of the SDSS DA
sample using various photometric systems and filter sets. The top row
in this figure is with the Pan-STARRS $grizy$ photometry. Although the
$gri$ histograms appear well-centered, the $y$ histogram in particular
shows a long, extended tail, indicating a potential problem with this
bandpass. By removing the $y$ magnitude from our fits, we obtain the
results displayed in the second row of Figure \ref{diff_mag_all} and
in the third panel of Figure \ref{comp_PHOT_all}. We notice that the
histograms using only the Pan-STARRS $griz$ photometry have not
changed significantly, with the exception of the $z$ histogram, which
now appears more centered. More importantly, however, the discrepancy
between the photometric temperatures in Figure \ref{comp_PHOT_all}
has remained essentially unchanged.

The third and fourth rows in Figure \ref{diff_mag_all} show our
results using SDSS $griz$ and $ugriz$ photometry, respectively. The
results using the full $ugriz$ photometric set are qualitatively
consistent with those shown in Figure 8 of \citet{Genest2014},
although the dispersion in their case is much smaller than ours, since
they restricted their DA sample to the best photometric fits (and
$\Te<20,000$~K), while ours is based on distance ($D<100$~pc). A more
interesting result is the comparison between Pan-STARRS and SDSS for
the same $griz$ filter sets (second and third row), which reveals a
much smaller dispersion with the Pan-STARRS photometry. This
demonstrates that the Pan-STARRS photometric system is superior to the
SDSS photometric system, {\it at least in a relative sense}, a
conclusion we also reached based on our inspection of the individual
photometric fits. Hence, despite the larger discrepancies between the
spectroscopic and photometric temperatures obtained from Pan-STARRS
$grizy$ photometry (as observed in Figure
\ref{comp_MASTERLIST_grizy}), it would be a shame not to take
advantage of its superior photometric quality.

\begin{figure*}[t]
\centering
\includegraphics[clip=true,trim=0cm 2cm 1cm 2cm,width=0.85\linewidth]{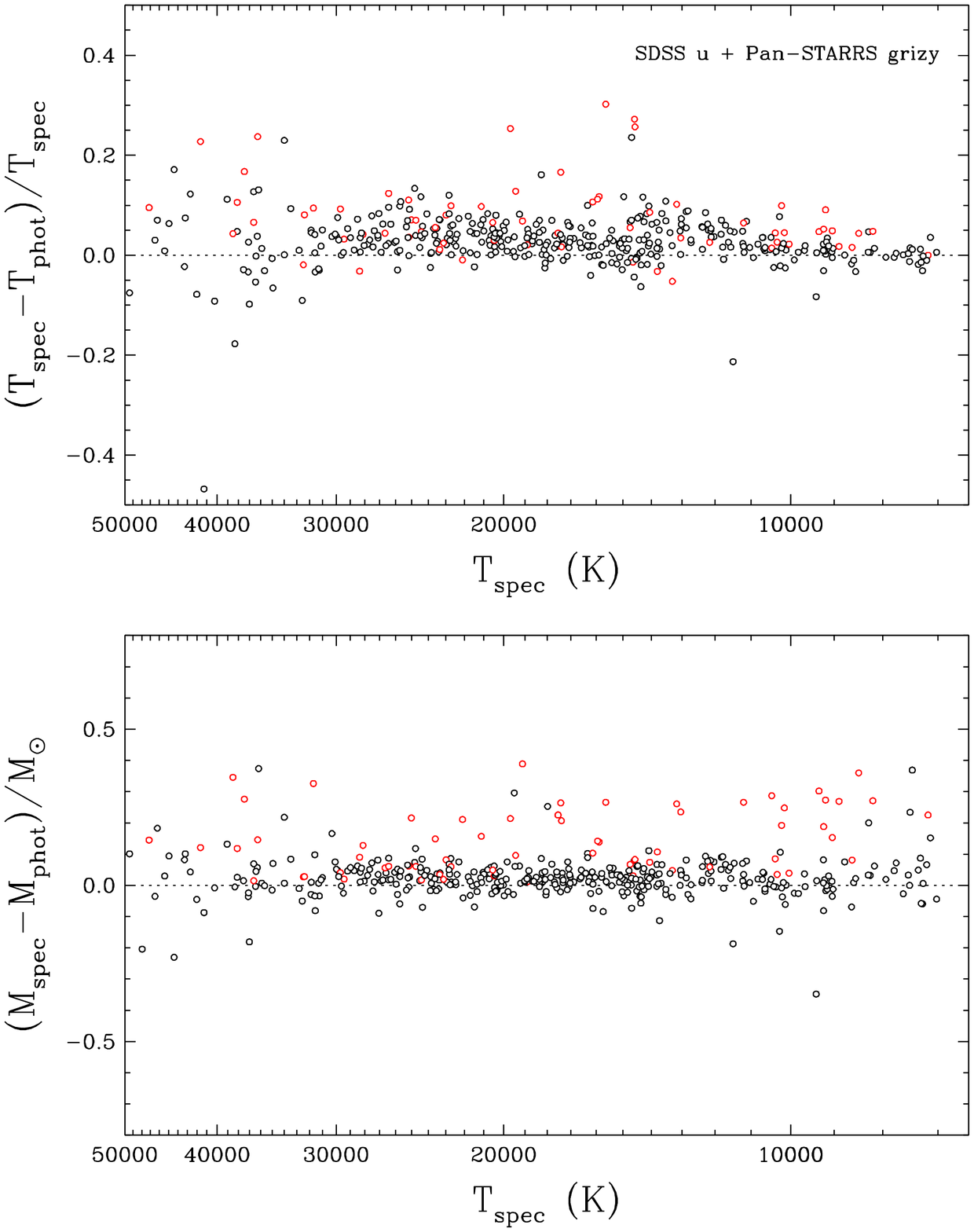}
\caption{Same as Figure \ref{comp_MASTERLIST_grizy} but by using
  photometric fits to the combined SDSS $u$ and Pan-STARRS $grizy$
  data.\label{comp_MASTERLIST_ugrizy}}
\end{figure*}

As a final experiment, we attempted to combine the SDSS $u$ bandpass
with the Pan-STARRS $grizy$ photometry, the results of which are
displayed in the bottom panels of Figures \ref{comp_PHOT_all} and
\ref{diff_mag_all}. Surprisingly, the resulting photometric
temperatures are now in excellent agreement with those obtained using
SDSS $ugriz$ photometry, which implies that the substitution of the
SDSS $griz$ with Pan-STARRS $grizy$ has no significant
effect on the measured photometric temperature, although the quality of our
fits has increased significantly. We adopted the same strategy for the
DA stars in the Gianninas et al.~sample, and included the SDSS $u$ magnitude,
whenever possible. The results are displayed in Figure
\ref{comp_MASTERLIST_ugrizy}. We see that the temperature discrepancy
has been significantly reduced, and that the photometric and
spectroscopic masses are now in much better agreement. These results
are now entirely consistent with those reported by GBB19 for the DA
stars in the SDSS. We mention again that we achieve results that are virtually
identical to those shown in Figure \ref{comp_MASTERLIST_ugrizy} if we
substitute the Pan-STARRS $grizy$ photometry with SDSS $griz$ data,
although individual fits are often superior with Pan-STARRS.
Unfortunately, SDSS $u$ magnitudes are available for only a fraction
of objects in our sample, a situation that will likely change, thanks
to the ongoing Canada-France Imaging Survey \citep{CFIS}, which will
map 10,000 square-degrees of the northern high Galactic latitude sky
in the $u$-band (CFIS-$u$).

\begin{figure}[t]
\centering
\includegraphics[clip=true,trim=1cm 1cm 1cm 0.5cm,width=\linewidth]{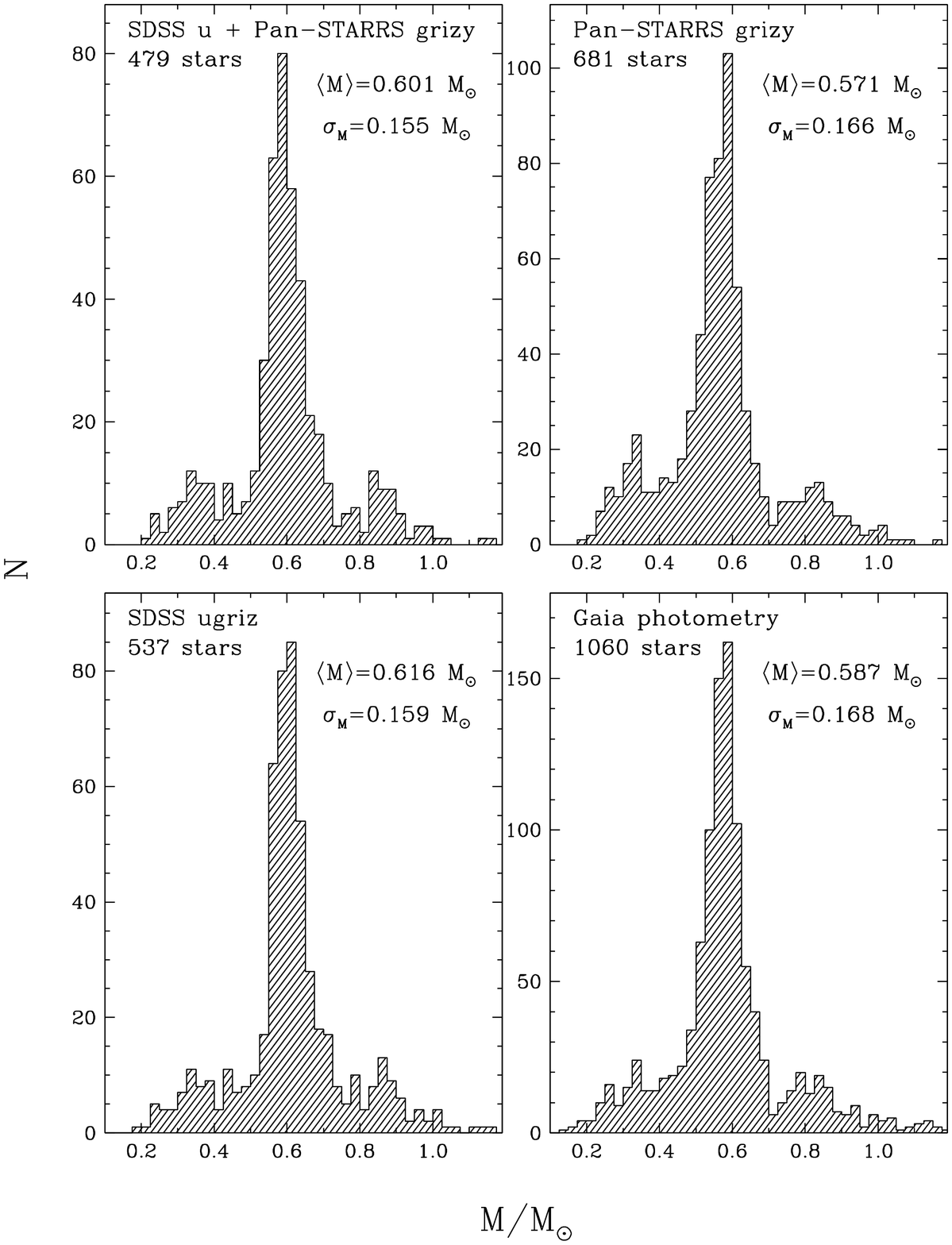}
\caption{Cumulative mass distributions measured using the photometric
  technique for DA white dwarfs drawn from the sample of
  \citet{Gianninas2011}.  Different photometric sets are used, as
  labeled in each panel, and the corresponding mean masses and
  dispersions are also given. Note that the number of stars,
    indicated in each panel, differs due to the availability of the
    photometry.\label{histo_mass_comp}}
\end{figure}

Finally, we also obtained results using {\it Gaia} photometry that are
similar to those shown in Figure \ref{comp_MASTERLIST_grizy}, in
particular the large offsets in temperature and mass. Instead of
repeating the same figure, we summarize our results in Figure
\ref{histo_mass_comp}, where the cumulative mass distributions are
displayed using four different sets of photometry: Pan-STARRS $grizy$,
SDSS $u$ + Pan-STARRS $grizy$ (corresponding to Figures
\ref{comp_MASTERLIST_grizy} and \ref{comp_MASTERLIST_ugrizy},
respectively), SDSS $ugriz$, and {\it Gaia} $G$, $G_{\rm BP}$, and
$G_{\rm RP}$ photometry. Note that the number of stars in each panel
is different due to the availability of the data in a given
photometric system. Also given in each panel are the average masses
and 1$\sigma$ dispersion values. As already mentioned above, the
results obtained with SDSS $u$ + Pan-STARRS $grizy$ and with SDSS
$ugriz$ are virtually identical, although the peak of the mass
distribution is more sharply defined using the Pan-STARRS photometry,
perhaps an indication of the superior quality of this particular data
set. If we omit the $u$ bandpass, however, we obtain average masses
that are lower, no matter whether we use Pan-STARRS $grizy$ alone or
{\it Gaia} photometry. Even though both of these subsamples are
significantly larger, nearly identical results are obtained if we
restrict our analysis to a common subsample.

The results displayed in Figure \ref{histo_mass_comp} can
probably explain the differences, of the same order ($\sim$0.03
\msun), between the average mass of DA white dwarfs in the SDSS
reported by \citet[][see their Figure 13]{Tremblay2019} based on {\it
  Gaia} photometry, $\langle M \rangle = 0.586$ \msun, and that found
by GBB19 (see their Figure 20) based on SDSS $ugriz$ photometry,
$\langle M \rangle = 0.617$ \msun. We also note that the mean mass
obtained by Tremblay et al.\ for DA stars in the Gianninas et
al.\ sample using {\it Gaia} photometry, $\langle M \rangle = 0.599$
\msun, is comparable to the value reported in Figure
\ref{histo_mass_comp}, $\langle M \rangle = 0.587$ \msun, even
though there are several differences between both analyses, the first
of which is the different procedure for taking into account
interstellar reddening. Also, our zero points for the {\it Gaia}
photometry are calculated under the assumption of zero magnitude for
Vega, which differ from their zero points listed in Table 3 of
\citet{GentileFusillo2019}.

Despite these differences, our conclusion remains the same. The
photometric temperatures obtained by neglecting the $u$ bandpass may be
significantly underestimated, especially for hotter stars in the
Rayleigh-Jeans regime. For a given luminosity, the photometric fit
will yield a larger radius, and thus a smaller mass, given the
mass-radius relation for white dwarfs. Consequently, the mean mass for
a given white dwarf sample will be smaller than the mean value obtained from fits
where the $u$ magnitude is included, by $\sim$0.03 \msun\ according to
our results.

\section{IMPLICATIONS FOR WHITE DWARF PHYSICS}\label{sec:implic}

\subsection{The Mass-Radius Relation Revisited}

In this section we revisit some of the results of \citet{Bedard2017}, who
presented a detailed spectroscopic and photometric analysis of 219 DA
and DB white dwarfs for which trigonometric parallax measurements were
available at that time, with the aim of testing the mass-radius
relation for white dwarfs. In order to compare physical quantities on
an equal footing, B\'edard et al.\ first compared the parallactic
distance, $D_\pi$, obtained directly from the trigonometric parallax,
to the distance inferred from the mass-radius relation, $D_{\rm MR}$.
The latter is calculated by first using evolutionary models to convert
the spectroscopic $\logg$ into radius, which is then combined with the
photometric value of $(R/D)^2$ to obtain the desired distance $D_{\rm
  MR}$.  The 1$\sigma$ confidence level between these two distance
estimates can then be calculated, including all sources of
uncertainties associated with both the spectroscopic and photometric
techniques.

\begin{figure*}[t]
\centering
\includegraphics[angle=270,clip=true, trim=5cm 2cm 5cm 2cm,width=\linewidth]{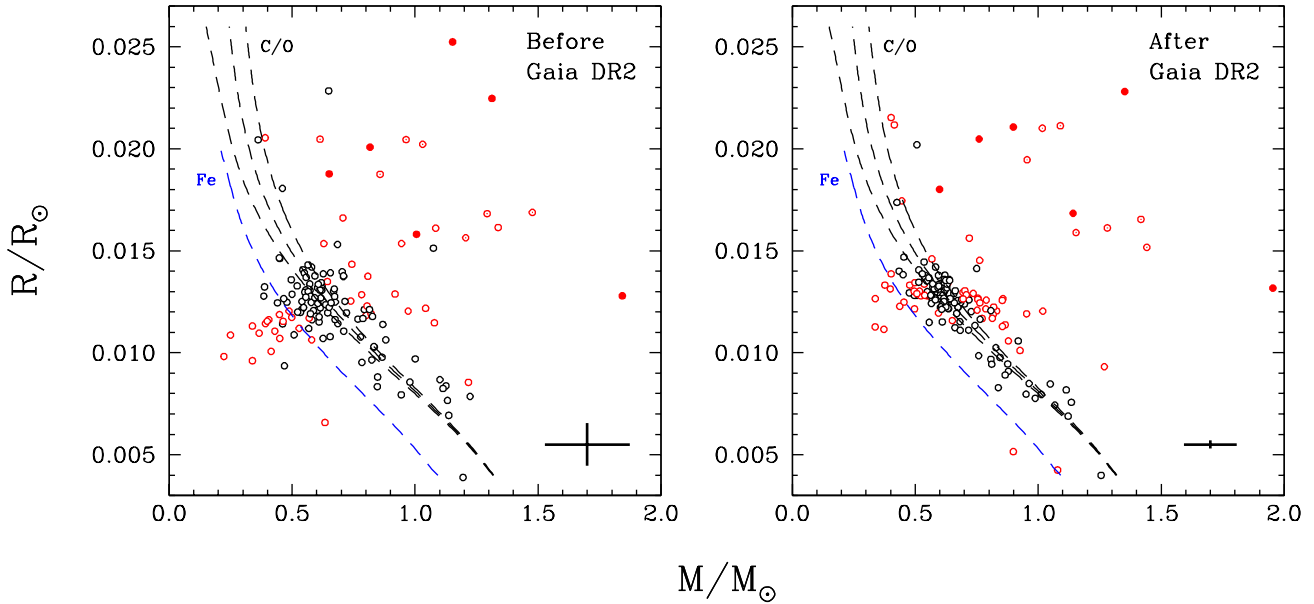}
\caption{Left panel: Location in the $R$ versus $M$ diagram for all
  white dwarfs in the sample analyzed by \citet{Bedard2017} with
  reliable trigonometric parallax and spectroscopic $\logg$
  measurements.  The stars shown in red exhibit differences larger
  than the 1$\sigma$ confidence level between $D_\pi$ and $D_{\rm MR}$
  (see text). The filled red circles and the dotted open circles
  correspond to known and suspected unresolved double degenerate
  systems, respectively. The cross in the lower right corner
  represents the 1$\sigma$ uncertainties averaged over the whole
  sample. Also shown are mass-radius relations for C/O-core, thick
  hydrogen envelope models at $\Te=7000$, 15,000, and 25,000~K (black
  dashed lines, from left to right), and for Fe-core, thick hydrogen
  envelope models at $\Te=15,000$~K (blue dashed line).  Right panel:
  same results but with the revised {\it Gaia} parallax measurements.
\label{compMR}}
\end{figure*}

The original results from B\'edard et al.~are reproduced in the left
panel of Figure \ref{compMR}, where all white dwarfs are plotted in
the $R$ versus $M$ diagram, together with various mass-radius
relations for different effective temperatures and core
compositions. Here, the radius $R$ is obtained directly from the
photometric technique, while the mass is derived by combining this
photometric radius with the spectroscopic $\log g$ ($g=GM/R^2$). The
right panel of Figure \ref{compMR} shows the same results but with the
updated trigonometric parallaxes from {\it Gaia}. With the exception
of the unresolved double degenerate binaries, or binary candidates
(shown as filled and dotted circles, respectively), we can see that
the white dwarfs are more closely packed near the theoretical
mass-radius relations with the {\it Gaia} parallaxes than with the
older measurements. However, we also note that the number of objects
with distances that differ by more than the 1$\sigma$ confidence level
(shown in red) has not changed significantly. This somewhat surprising
result is due to the fact that the errors on the {\it Gaia} parallaxes
are now extremely small --- as can be estimated by the average
  1$\sigma$ uncertainty on the radius $R$ displayed in both panels of
Figure \ref{compMR} --- while the errors on the mass are still
dominated by the large uncertainties of the spectroscopic $\logg$
determinations.

Several puzzling discrepancies reported by B\'edard et al.~are worth
discussing here. The first case is G87-7, whose distance obtained from
C/O-core models, $D_{\rm MR}=17.5$ pc, was found to be significantly
different from the parallactic distance, $D_{\pi}=15.7$ pc, measured
by {\it Hipparcos}. B\'edard et al.~suggested that G87-7 could have
instead an iron core, since the distance inferred from Fe-core models,
$D_{\rm MR}=15.9$ pc, was in much better agreement with the
parallactic distance. This discrepancy has been resolved with the {\it
  Gaia} parallax, $D_\pi=17.07$ pc, which is now in excellent
agreement with the distance inferred from C/O-core models. Two
additional objects of interest are the ZZ Ceti white dwarfs Ross 548
and GD 1212, for which B\'edard et al.\ obtained from C/O-core models,
$D_{\rm MR}=32.6$ pc and 18.7 pc, respectively, in strong disagreement
with the parallactic distances of $D_\pi=63.3$ pc and 15.9 pc,
respectively. In both cases, no physical model could explain the observed
discrepancies. Fortunately, the improved parallax measurements for the
two ZZ Ceti stars, respectively $D_\pi=32.76$ pc and 18.66 pc, are now
in perfect agreement with the distances inferred from C/O-core models.

We end this section by discussing another puzzling result. In the
analysis of \citet{Bedard2017}, the agreement between their
spectroscopic and photometric parameters was found to be excellent,
especially for effective temperatures (see their Figure 4), in sharp
contrast with the results discussed in GBB19 and those reported here,
where differences of the order of 10\% are found between the two
temperature estimates. However, if we plot {\it differences} between
temperatures rather than comparing values against each other, we do
find a small systematic offset. Also, the sample of B\'edard et
al.~contains very few hot stars, and if we superpose their results
with those shown in Figure \ref{comp_MASTERLIST_ugrizy} for the complete
Gianninas et al.~sample, the temperature discrepancies observed in
both samples are entirely consistent.

\subsection{Mass-Temperature Distributions}

\subsubsection{The Sample of Spectroscopically Identified White Dwarfs}

We now investigate further the physical parameters obtained from
photometric fits by first looking at the white dwarfs identified
spectroscopically in the MWDD. Given our results above, we combine the
SDSS $u$ magnitude with the Pan-STARRS $grizy$ photometry since this
sample is completely dominated by objects identified in the SDSS (over
95\% of the objects in our sample include the SDSS $u$ magnitude).  We
rely solely on Pan-STARRS photometry if the SDSS $u$ bandpass is not
available. Again, we restrict our analysis to {\it Gaia} parallaxes
with uncertainties smaller than 10\%.

\begin{figure*}[t]
\centering
\includegraphics[angle=270,clip=true,trim=3.5cm 2cm 4cm 2cm,width=\linewidth]{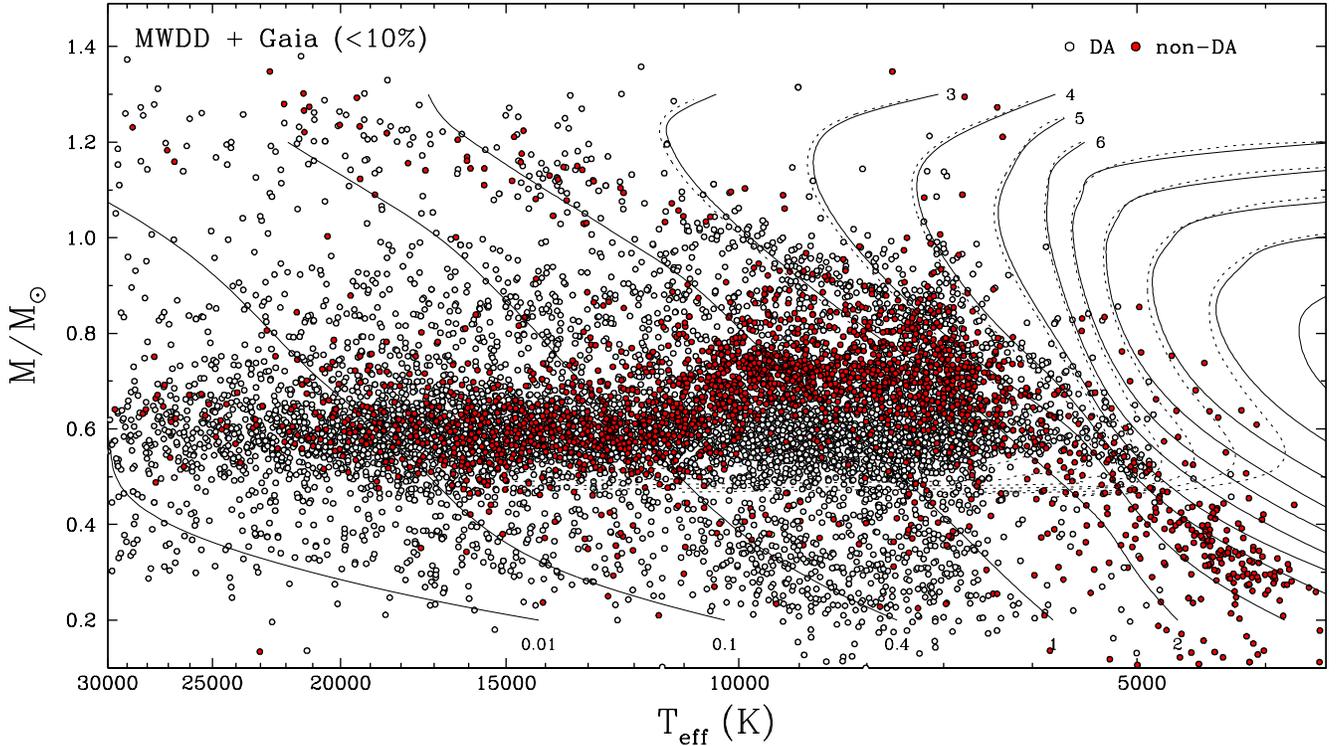}

\caption{Stellar masses as a function of effective temperature for all
  spectroscopically identified white dwarfs in the MWDD. The
  parameters have been determined using photometric fits to SDSS $u$
  and Pan-STARRS $grizy$ photometry, mostly, and {\it Gaia} parallaxes
  with $\sigma_\pi/\pi<0.1$. DA (white symbols, 7340 objects) and
  non-DA (red symbols, 2348 objects) white dwarfs have been fitted
  under the assumption of pure hydrogen and pure helium models,
  respectively. Also shown as solid curves are theoretical isochrones,
  labeled in units of $10^9$ years, obtained from cooling sequences
  with C/O-core compositions, $q({\rm He})=10^{-2}$, and $q({\rm
    H})=10^{-4}$, while the dotted curves correspond to isochrones
  with the main sequence lifetime taken into
  account.\label{correltm_MWDD_pureHe}}
\end{figure*}

Figure \ref{correltm_MWDD_pureHe} shows the stellar masses
as a function of $\Te$ for all DA and non-DA white dwarfs identified
spectroscopically in the MWDD. Here we simply assume pure hydrogen and
pure helium compositions for the DA and non-DA stars, respectively,
and we restrict our analysis to a range of effective temperatures
where the photometric technique is the most reliable ($\Te<30,000$~K,
GBB19). The stellar masses have been derived from the measured stellar
radii using evolutionary models\footnote{See
  http://www.astro.umontreal.ca/$\sim$bergeron/CoolingModels.} similar
to those described in \citet{Fontaine2001} with (50/50) C/O-core compositions,
$q({\rm He})\equiv M_{\rm He}/M_{\star}=10^{-2}$, and $q({\rm
  H})=10^{-4}$ or $10^{-10}$ for DA and non-DA stars,
respectively. Also shown are theoretical isochrones, with and without
the main sequence lifetime taken into account, as described in
\citet[][see their Section 5.5]{BLR01}.

As mentioned above, the white dwarfs displayed in Figure
\ref{correltm_MWDD_pureHe} are completely dominated by those
identified in the SDSS, and consequently, there is a strong bias towards
hotter objects given the color cuts inherent to this survey.  If we
first consider the DA stars, we can see that the mass distribution is
well centered around 0.6 \msun, with a significant number of high-mass
and low-mass white dwarfs. The low-mass DA stars ($M\lesssim0.5$
\msun) in this diagram correspond most certainly to unresolved double
degenerate binaries whose photometric masses are meaningless since
these have been obtained under the assumption of a single star. The
high-mass DA stars, on the other hand, are believed to be the end
result of stellar mergers \citep{Kilic2018}, or alternatively, they
can also be explained as the outcome of the initial-to-final mass
relation \citep{ElBadry2018}. Note that the results presented here for
the DA stars are entirely consistent with those displayed in Figure 7
of GBB19.

The situation is relatively more complex for the non-DA white dwarfs
shown in Figure \ref{correltm_MWDD_pureHe}. As a reminder, these
non-DA white dwarfs include DB, DQ, DZ, DC, and all other subtypes (DO
stars are outside the temperature range displayed here). Below
$\Te\sim5000$~K, the DC spectral type gives no indication about the
atmospheric composition since both hydrogen and helium lines become
invisible at these temperatures (see, e.g., \citealt{BRL97}). The DC
stars in this temperature range may as well have hydrogen-dominated
compositions. We defer our discussion of these objects further
below. Above $\Te\sim11,000$~K, the non-DA population is dominated by
DB/DBA white dwarfs, whose masses are well centered around 0.6 \msun,
in agreement with the results of GBB19 (see their Figure 8). We also
see that the number of extremely high- and extremely low-mass non-DA
white dwarfs is relatively small compared to DA stars. Actually, the
high-mass, non-DA stars above $\sim$10,000~K observed here correspond
in majority to warm DQ white dwarfs (Coutu et al. 2019, in
preparation).

The most striking feature in Figure \ref{correltm_MWDD_pureHe}, by
far, is the large number of non-DA white dwarfs below
$\Te\sim11,000$~K with masses significantly above 0.6~\msun. The
number of non-DA stars in this temperature range with masses around
0.6~\msun\ is actually quite small. The ``step function'' in mass
observed here corresponds exactly to the discontinuity in the
color-magnitude diagram described in Figure \ref{color_mag_mass}.
Such high masses inferred from photometric fits is reminiscent of a
similar problem observed in the context of cool DQ stars analyzed with
pure helium models. For instance, \citet[][see their Figure
  8]{Dufour05} compared the effective temperatures and masses derived
from pure helium models with those obtained with models including
carbon. The temperatures measured from models including carbon are
found to be significantly lower than the pure helium solutions. In
this case, the inclusion of carbon in the equation of state increases
the number of free electrons and thus the contribution of the He$^-$
free-free opacity, which in turn affects the atmospheric structure, in
particular in the continuum forming region. Since the derived
temperatures are significantly reduced, larger stellar radii --- and
thus smaller masses --- are required to match the observed stellar
flux.  In some cases, the masses are decreased by as much as 0.2
\msun\ when carbon is included.

The key ingredient in the above argumentation is the presence of
additional free electrons in otherwise pure helium atmospheres,
whether these are coming from carbon, other metals, or even hydrogen.
Note that in this context, GBB19 have demonstrated (see their Figure
6) that the presence of hydrogen in DBA white dwarfs does not affect
the masses inferred from photometry in the temperature range where DB
stars are found ($\Te\gtrsim 11,000$~K). Hence, their measured masses
in Figure \ref{correltm_MWDD_pureHe} are free of this uncertainty and
completely reliable. In fact, they overlap perfectly with those of DA
stars in the same temperature range. But the almost complete absence
of normal mass ($M\sim0.6$ \msun) non-DA stars below 11,000~K suggests
that pure helium white dwarfs are extremely rare, if they exist at
all. In the range $11,000\ {\rm K}\lesssim\Te\lesssim6000\ {\rm K}$,
we actually find that about 25\% of the objects are DQ stars, while
another 25\% are DZ stars, but most are DC white dwarfs.

It is a well known fact the ratio of non-DA to DA white dwarfs
increases dramatically below $\Te\sim10,000$~K (see, e.g.,
\citealt{FW1987}). The sudden increase in the number of non-DA stars
in this temperature range has been interpreted as the result of the
mixing of the superficial convective hydrogen layer with the deeper
and much more massive convective helium envelope (see
\citealt{Rolland2018} and references therein). In this mixing
scenario, the relatively small amount of hydrogen in the upper layers
is being thoroughly mixed within the underlying helium convection zone,
resulting in photospheric hydrogen abundances that are extremely small,
but not zero. \citet{Rolland2018} have explored this scenario more
quantitatively by using state-of-the art envelope models to predict
the hydrogen-to-helium abundance ratio as a function of the
temperature at which convective mixing occurs (see their Figure
16). Qualitatively, the thicker the hydrogen layer, the cooler the
mixing temperature, and the larger the predicted photospheric hydrogen
abundance upon mixing. After mixing occurs, the hydrogen abundance
remains almost constant with time. In some cases, the presence of
hydrogen might be revealed by the detection of a weak H$\alpha$
absorption feature, heavily broadened by van der Waals interactions,
as observed for instance in L745-46A and Ross 640 (see Figure 14 of
\citealt{Giammichele2012}); other examples from the SDSS can be found in
\citet{Rolland2018}. But as the star cools off, the weak H$\alpha$
feature rapidly falls below the limit of visibility. Hence in more
representative cases, the object becomes a DC white dwarf, or a DZ star if
the atmosphere was already contaminated with metals.

\begin{figure*}[t]
\centering
\includegraphics[angle=270,clip=true,trim=3.5cm 2cm 4cm 2cm,width=\linewidth]{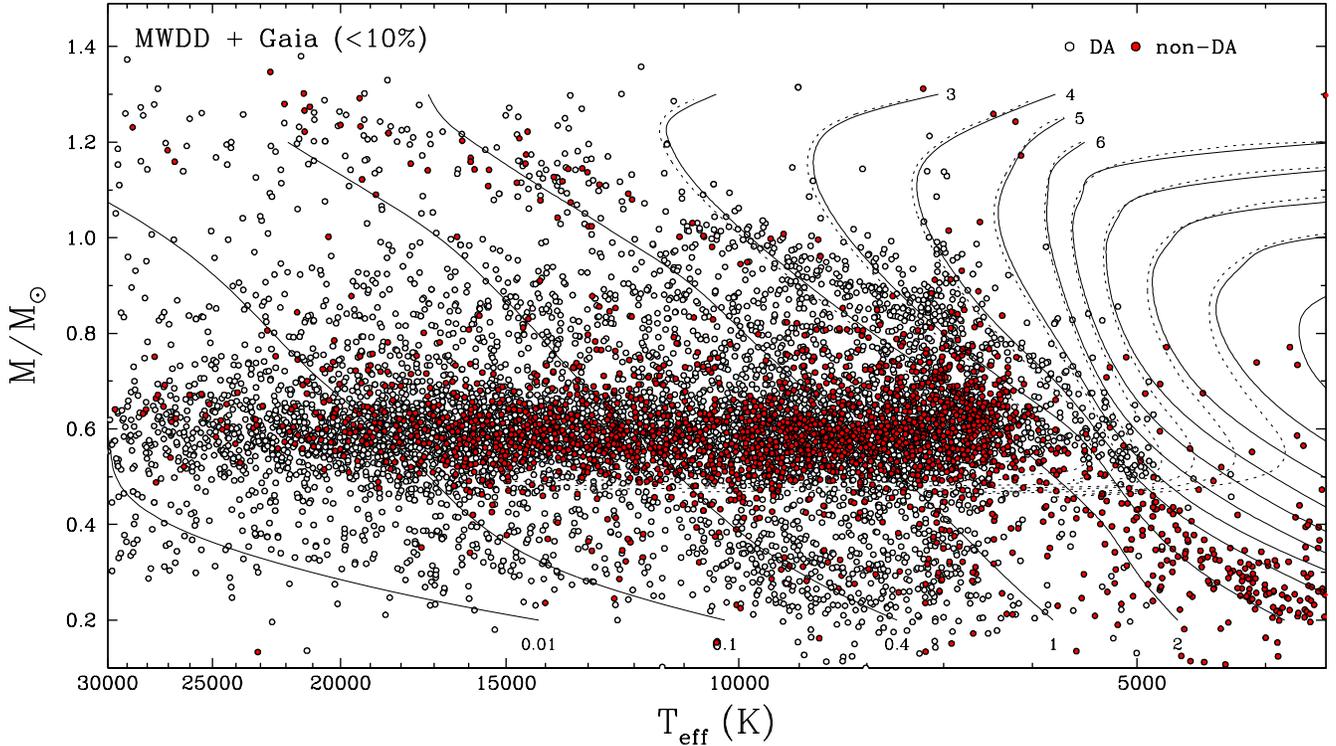}
\caption{Same as Figure \ref{correltm_MWDD_pureHe} except
  that the non-DA white dwarfs have been fitted under the assumption
  of mixed H/He models (see text).\label{correltm_MWDD_mixed}}
\end{figure*}

With this idea in mind, we performed the following experiment. Instead
of fitting the non-DA stars in our sample with pure helium models, we
used mixed H/He models where the hydrogen abundance is adjusted as a
function of effective temperature. In the range of DB stars
($\Te>11,000$~K), we adopt a typical value of $\log\nh=-5$, while
below this temperature, we gradually increase the hydrogen abundance
following the predictions of the convective mixing scenario displayed
in Figure 16 of \citet{Rolland2018}. Below 6000~K, where these
simulations stop and where we have very few objects in our sample, we
assume the same composition as in DB white dwarfs. As mentioned above,
the cool non-DA stars in this temperature range may have
hydrogen-dominated compositions, and these will be discussed
separately below. The results of our experiment are displayed in
Figure \ref{correltm_MWDD_mixed}.  As anticipated, the masses for the
DB stars have not changed since in this temperature range, helium
remains the main electron donor.  At lower temperatures, however, the
large masses observed in Figure \ref{correltm_MWDD_pureHe} have been
significantly reduced, and now overlap perfectly with those of DA
stars. Our results clearly indicate that the photometric parameters
obtained from pure helium models may be unreliable, and that more
reasonable masses are derived when additional electron donors are
included, whether they are in the form of metals (including carbon) or
hydrogen.  We stress here that this is only an experiment, and that
stellar parameters of individual white dwarfs can only be obtained
through a tailored analysis of each object with an appropriate model
atmosphere grid. In that sense, accurate parameters for DC stars may
be impossible to achieve since only upper limits on the hydrogen
abundance can be measured. One thing our results indicate, however, is
that this limit is certainly not zero.

\subsubsection{The Sample of White Dwarfs in {\it Gaia}}\label{sec:gaiawd}

\begin{figure*}[t]
\centering
\includegraphics[clip=true,trim=0cm 1cm 2cm 1cm,width=0.7\linewidth]{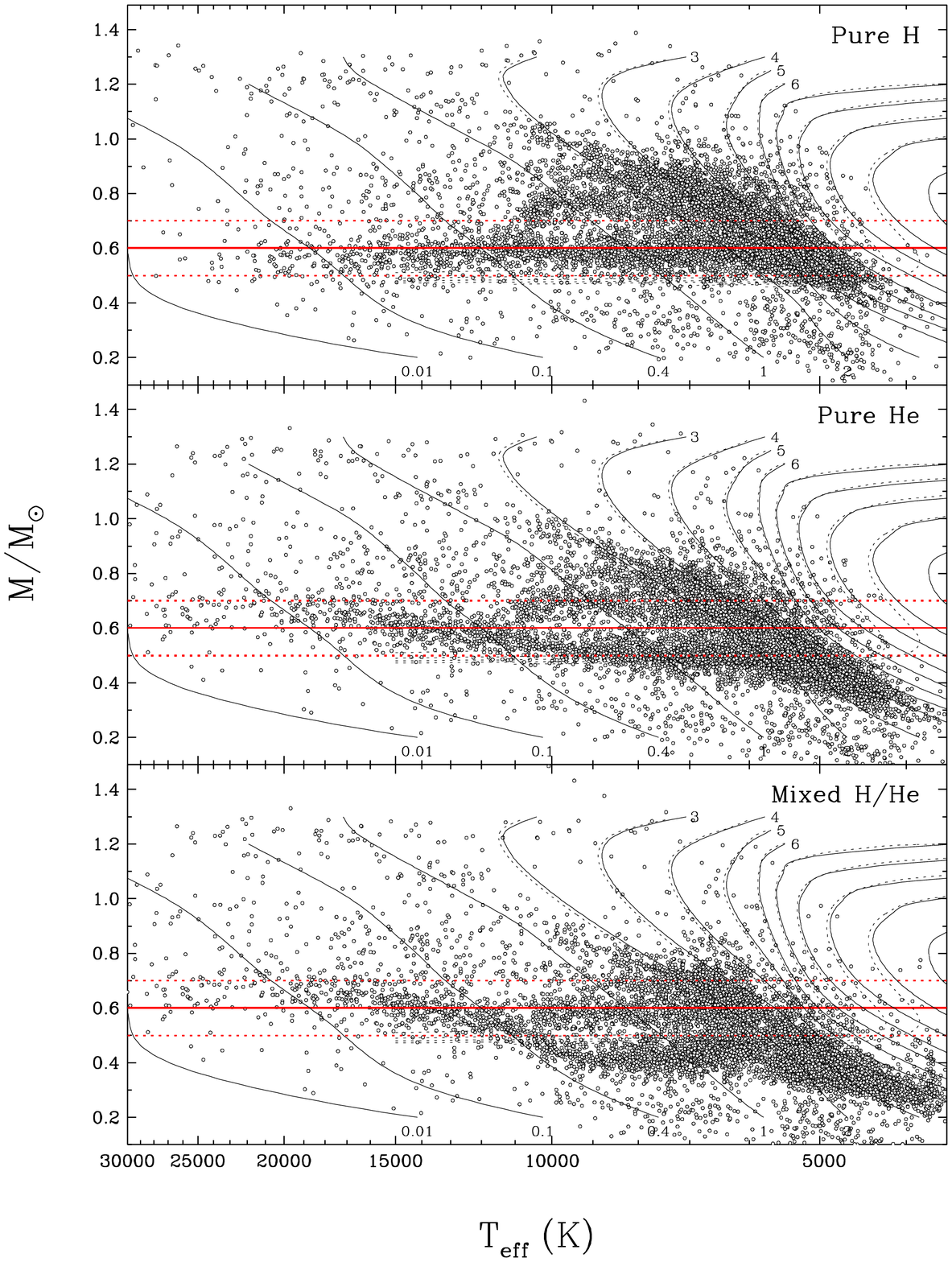}
\caption{Stellar masses as a function of effective temperature for all
  white dwarfs found in {\it Gaia} within a distance of 100 pc,
  including both spectroscopically identified white dwarfs as well as
  white dwarf candidates. The parameters have been determined using
  photometric fits to Pan-STARRS $grizy$ photometry and {\it Gaia}
  parallaxes with $\sigma_\pi/\pi<0.1$. The white dwarfs in each panel
  have been fitted under the assumption of pure hydrogen, pure helium,
  or mixed H/He models, as indicated in the figure. Also shown are the
  same isochrones as in Figure
  \ref{correltm_MWDD_pureHe}. The horizontal lines represent
  constant masses of $M=0.6\pm0.1$
  \msun.\label{correltm_PanSTARRS_100pc}}
\end{figure*}

We now turn our attention to white dwarfs found in {\it Gaia}, which
include both spectroscopically identified white dwarfs as well as
white dwarf candidates. We restrict our analysis to objects with a
distance $D<100$~pc for which interstellar reddening is assumed to be
negligible. If we had access to the $u$ bandpass for all objects, we
could in principle differentiate hydrogen-rich from helium-rich
objects in the temperature range where the Balmer jump is important
(see Figure \ref{color_color}). Since $u$ magnitudes are available for
only a limited number of objects in our sample, we adopt a simpler
approach and fit only the Pan-STARRS $grizy$ photometry by assuming
that all white dwarfs have either pure hydrogen, pure helium, or
mixed H/He atmospheric compositions (as defined above).  Also, because
we do not include the $u$ bandpass, our stellar masses for hot stars
might be underestimated by $\sim$0.03 \msun\ on average.  The stellar
masses obtained under these assumptions are displayed in Figure
\ref{correltm_PanSTARRS_100pc} as a function of effective temperature.

The distribution of {\it Gaia} white dwarfs within 100 pc shown in
Figure \ref{correltm_PanSTARRS_100pc} is markedly different from that
displayed in Figure \ref{correltm_MWDD_pureHe} (or
\ref{correltm_MWDD_mixed}), regardless of the assumed atmospheric
composition, since the latter is mostly based on UV-excess selection
criteria rather than distances.  Consequently, the number of cool
white dwarfs in our distance-limited sample is significantly larger,
while the number at higher temperatures is much smaller. At high
temperatures ($\Te\gtrsim11,000$~K), the differences in mass inferred
from pure H or pure He models --- applicable to DA and DB stars,
respectively --- are of the order of $\sim$0.1 \msun, but the
differences between the pure He and mixed H/He solutions in the same
temperature range are completely negligible, in agreement with our
discussion above. At lower temperatures, however, the effects due to
atmospheric composition are much more pronounced. Somewhat
unexpectedly, the average mass drops as we go from pure H to pure He,
and then to mixed H/He models. In particular, the bulk of massive
white dwarfs in the $6000-10,000$~K temperature range has an average
mass above 0.7 \msun\ under the assumption of pure hydrogen
atmospheres, and of $\sim$0.6 \msun\ with mixed H/He models.

\begin{figure}[t]
\centering
\includegraphics[clip=true,trim=3cm 4.5cm 3.5cm 4cm,width=\linewidth]{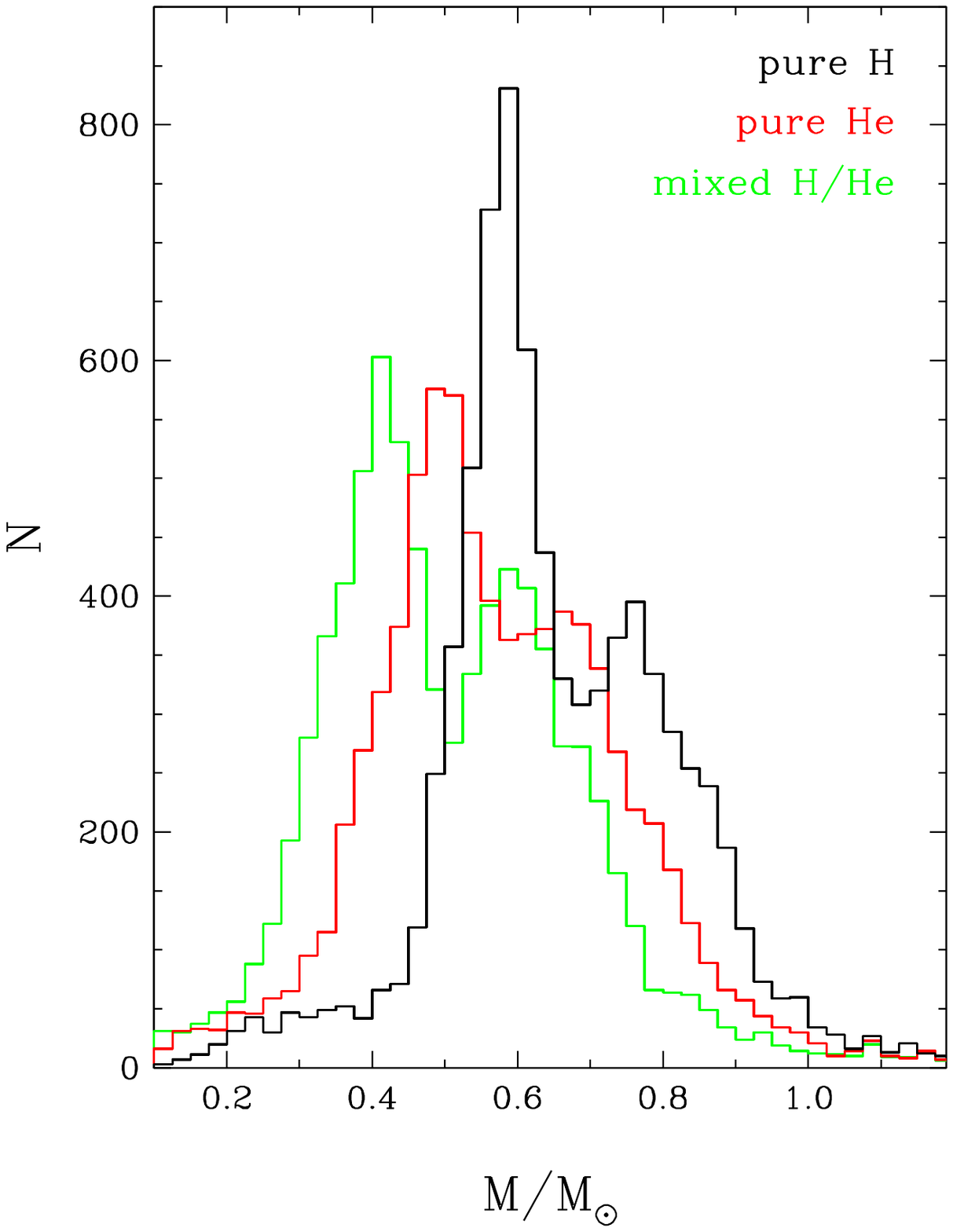}
\caption{Cumulative mass distributions for all white dwarfs found in
  {\it Gaia} within a distance of 100 pc (also displayed in Figure
  \ref{correltm_PanSTARRS_100pc}), including both spectroscopically
  identified white dwarfs as well as white dwarf candidates.  All
  objects in each mass distribution have been fitted under the
  assumption of pure hydrogen, pure helium, or mixed H/He models, as
  indicated in the figure.
\label{histo_mass_100pc}}
\end{figure}

We can explore these results more quantitatively by looking at the
cumulative mass distributions derived with our various assumptions
about the atmospheric composition, the results of which are displayed
in Figure \ref{histo_mass_100pc}. The pure hydrogen models yield the
sharpest peak around 0.6 \msun, but also a well-developed high-mass
bump with a peak around 0.8 \msun. This mass distribution is
completely analogous to that reported by \citet[][see their Figure
  5]{Kilic2018}; in this case, the high-mass bump has been interpreted
by the authors as evidence of a large population of merged white
dwarfs. We can see, however, that this high-mass bump virtually
vanishes if we use mixed H/He models.  Pure helium models yield
results that are in between. Obviously, the true mass distribution
requires the knowledge of the atmospheric composition of each
individual white dwarf in our sample. Nevertheless, our results
indicate that the number of massive ($M\sim0.8$ \msun) white dwarfs is
certainly much smaller than previously reported by Kilic et al.

Another feature worth discussing in Figure
\ref{correltm_PanSTARRS_100pc} is the behavior at the cool end of the
distribution ($\Te\lesssim5000$~K). Both the pure helium and the mixed
H/He models give rise to a large number of extremely low-mass white
dwarfs --- $M< 0.5$ \msun, and even $M<0.4$ \msun\ with mixed
models. On the other hand, pure hydrogen models bring the masses of
most cool white dwarfs comfortably above 0.5 \msun. Since pure
hydrogen white dwarfs in this temperature range are featureless, we
predict that most, but not all, cool white dwarfs identified by {\it
  Gaia} are H-rich DC stars.

\subsection{The Crystallization Sequence}\label{sec:crys}

\begin{figure*}[t]
\centering
\includegraphics[angle=270,clip=true,trim=3.5cm 2cm 4cm 2cm,width=\linewidth]{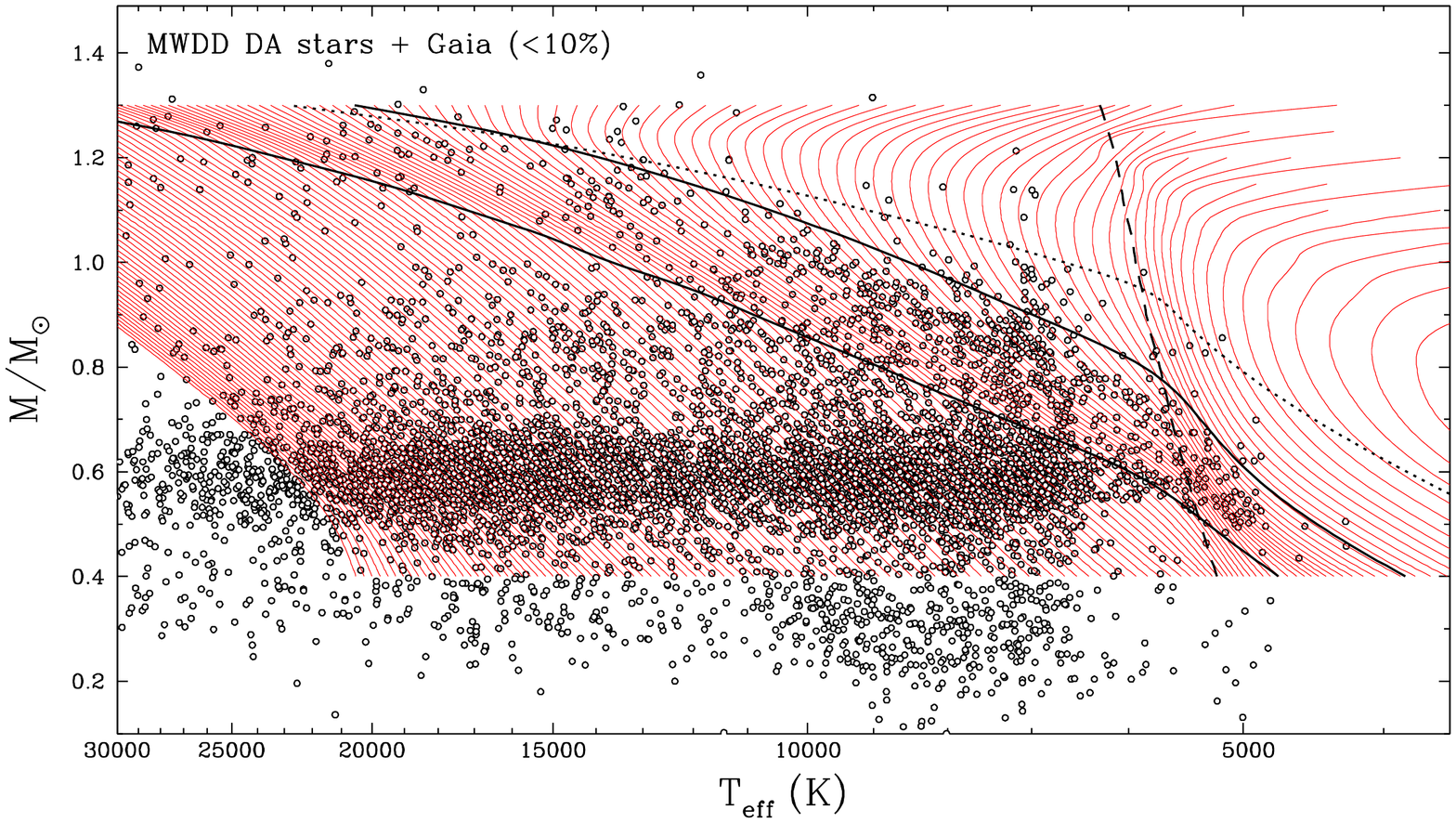}
\caption{Stellar masses as a function of effective temperature for all
  DA white dwarfs spectroscopically identified in the MWDD. The
  parameters have been determined using photometric fits to SDSS $u$
  and Pan-STARRS $grizy$ photometry, and {\it Gaia} parallaxes with
  $\sigma_\pi/\pi<0.1$. Also shown are theoretical isochrones (white
  dwarf cooling time only) obtained from cooling sequences with
  C/O-core compositions, $q({\rm He})=10^{-2}$, and $q({\rm
    H})=10^{-4}$, equally spaced by $\Delta\log\tau_{\rm cool}=0.02$
  (in years). The lower solid curve indicates the onset of
  crystallization at the center of evolving models, while the upper
  one indicates the locations where 80\% of the total mass has
  solidified. The dotted curve corresponds to the transition between
  the classical to the quantum regime in the ionic plasma, and the
  dashed curve indicates the onset of convective
  coupling.\label{crystallization}}
\end{figure*}

It is instructive to reconsider the mass-effective temperature
distribution of {\it Gaia} DA white dwarfs in the light of the recent
discovery of the signature of crystallization --- a characteristic
pile-up forming a sequence going across evolutionary tracks --- in the
{\it Gaia} color-magnitude diagram for such stars
\citep{TremblayNat2019}.  To this end, we plotted in Figure
\ref{crystallization} our most reliable estimates of mass and
effective temperature for spectroscopically identified DA white dwarfs
having {\it Gaia} parallaxes with $\sigma_\pi/\pi<0.1$. Next, using
the DA subset of the standard evolutionary models which we briefly
described above in Section 4.1 (these are the same models used also in
\citealt{TremblayNat2019}), we plotted a series of tightly-spaced
cooling isochrones in the mass-effective temperature plane. The
(variable) density of these many isochrones indicates graphically
phases of slowing down and of accelerated cooling. Note that our
(50/50) C/O core models include the release of latent heat upon
crystallization, but no additional source of energy associated with
possible phase separation between C and O. Further details concerning
our approach to the crystallization process can be found in
\citet{TremblayNat2019}.

The lower solid curve in Figure \ref{crystallization} corresponds to
the onset of crystallization at the center of an evolving model (at
constant mass, from left to right). From that point on, with further
cooling, the solidification front progresses upward in the star from
the center, and latent heat is progressively released. By the time
some 80\% of the total mass of the star has solidified --- this is
indicated by the upper solid curve --- most of the latent heat has
been spent. The release of latent heat corresponds to a slowdown in
the evolution of a white dwarf (specifically, the cooling rate
decreases), and this is well illustrated in the tightening up of the
isochrones in between the two solid curves.  Note that the DA white
dwarfs found in between the two solid curves in Figure
\ref{crystallization} correspond to a subset of the stars populating
the ``crystallization sequence'' reported by \citet{TremblayNat2019}
in their observational {\it Gaia} color-magnitude diagram (see their
Figure 2). They are less numerous, but they do define a pile-up
structure that is clearly associated with the release of latent heat.

The most significant effect of crystallization on the evolution of
white dwarfs, however, is not the slowdown associated with the release
of latent heat, but rather the so-called Debye cooling phase, i.e.,
the subsequent transition, in the solid phase, from the classical
regime where the specific heat of a solid is independent of
temperature to the quantum regime where it goes down from that
constant value with decreasing temperature. In the quantum regime, the
specific heat decreases quickly with cooling, which depletes rapidly
the reservoir of thermal energy and produces a spectacular increase of
the cooling rate, leading to the concomitant rapid shift to the black
dwarf phase. This is well illustrated in Figure \ref{crystallization},
through the turnover of the isochrones toward low effective
temperatures, most obviously seen in the more massive models. For the
present purposes, we loosely defined the transition from the classical
to the quantum regime through the dotted curve. The latter has been
obtained by isolating the evolving model where the central temperature
becomes, from above, equal to the central Debye temperature (function
of density and composition, but not of temperature). According to our
models, a few massive DA white dwarfs in our sample --- those above
the dotted curve --- are likely crystallized white dwarfs having
reached the quantum regime and undergone rapid Debye cooling. They
must be still quite young ($\tau_{\rm cool}\le 4.5$ Gyr; see Figure
\ref{correltm_MWDD_pureHe}) and their predecessors must already
have faded away to the black dwarf state.

Beyond crystallization, another phenomenon complicates things in the
late cooling history of white dwarfs, and that is convective coupling
(see, e.g., \citealt{Tremblay2015}). The onset of convective coupling,
when superficial convection first reaches into the degenerate core
(where resides essentially all of the thermal energy), is indicated by
the dashed curve in Figure \ref{crystallization}. The phenomenon is
weakly mass-dependent, and clearly interacts with the manifestations
of crystallization as can be observed through the changes of slope in
the solid and dotted curves to the right of the convective coupling
boundary. For a middle-of-the-road DA white dwarf with $\sim$0.6
\msun, crystallization and convective coupling occur at nearly the
same time, so it is almost impossible to untangle their effects. Only
for the larger masses are the two mechanisms occurring during
different phases of their evolution, and this is well depicted in
Figure \ref{crystallization}. For example, picking the case of a 1.0
\msun\ model, it is clear from the figure that the star is already
highly solidified and in its Debye phase by the time convective
coupling turns on. When it does, it acts in two phases (like
crystallization), and this is well illustrated by the behavior of the
isochrones to the right of the dashed curve for that particular
mass. First, there is an initial release of thermal energy as the
outer envelope becomes much more transparent to photons through
convective transport (the isochrones bunch together), and, second,
this is followed by a phase of accelerated cooling, convective cooling
(the isochrones separate from each other), compared to the purely
radiative case (see \citealt{Tremblay2015} for details). Note that
convective cooling, in the presence of Debye cooling as is the case in
this example, accelerates considerably the passage to the black dwarf
state.

For more representative masses, the effects of crystallization and
convective coupling manifest themselves initially in their phases of
slowing down which add up together (release of latent heat superposed on
the extra thermal energy liberated when the envelope becomes suddenly
more transparent through convection). This is best seen in the behavior
of the isochrones which come very close to each other, just to the right
of the dashed curves and in between the two solid curves in the range
0.5 to 0.6 \msun\ . This is where a maximum pile-up of stars is
expected. Unfortunately here, our sample of DA white dwarfs is limited
by the availability of suitable photometric data, and one cannot test
for this possibility. However, we consider a larger sample in what
follows.

\subsection{Constraint on the Core Composition of {\it Gaia} White
  Dwarfs}\label{sec:core} 

\begin{figure*}[t]
\centering
\includegraphics[clip=true,trim=1cm 3.5cm 1cm 3.5cm,width=0.85\linewidth]{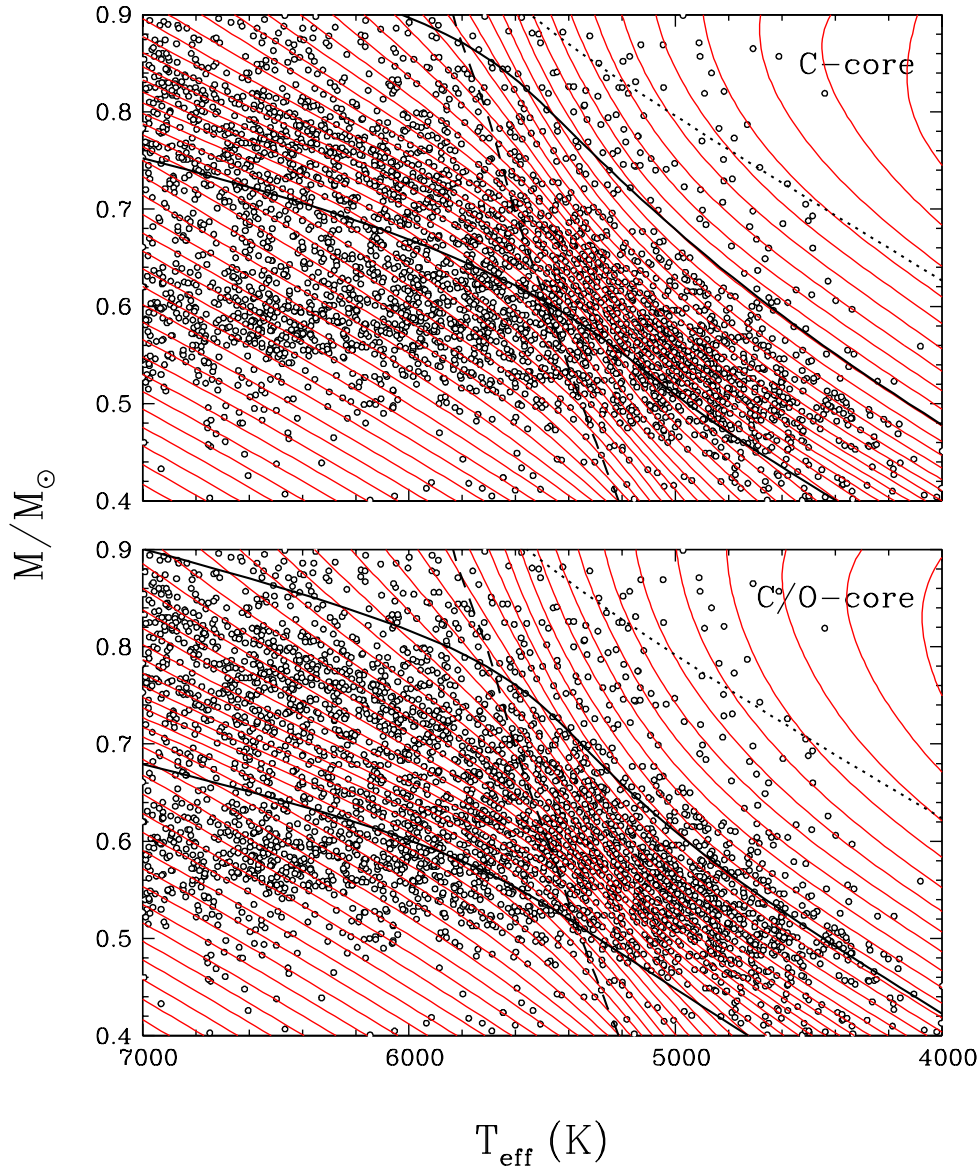}
\caption{Magnified view of the low-mass, low-effective temperature
  corner of the upper panel of Figure \ref{correltm_PanSTARRS_100pc}
  showing the highest-density clump of stars observed in the empirical
  data. The curves are similar to those found in Figure
  \ref{crystallization}. The lower (upper) panel refers to C/O-core
  (pure C-core) models. Note that, unlike the curves defining the
  convective coupling and the classical-to-quantum boundaries, the
  locations of the isochrones and of the theoretical crystallization 
  sequence (the two solid curves) depend strongly on the core
  composition of the models. \label{crystallization_100pc_zoom}}  
\end{figure*}

We reconsider the sample of all white dwarfs found in {\it Gaia} within a
distance of 100 pc, including both spectroscopically identified white
dwarfs as well as white dwarf candidates. Both panels of 
Figure \ref{crystallization_100pc_zoom} are a zoomed-in view of the
mass-effective temperature distribution depicted in the upper panel of
Figure \ref{correltm_PanSTARRS_100pc}, the latter obtained under the specific
assumption that the atmospheres of all these stars are made of pure
hydrogen. We emphasize, once again, that detailed atmospheric analyses
are required on a case-by-case basis, but our approach has merits from a
statistical standpoint, especially in view of our suggestion above that
the majority of the coolest white dwarfs in the {\it Gaia} sample likely
have hydrogen-dominated atmospheres.

In the lower panel of Figure \ref{crystallization_100pc_zoom}, one's
attention is attracted by the strong correlation that exists between the
behavior of the isochrones that bunch closest together and the maximum
density of white dwarfs observed in an area roughly centered around
$\Te\sim 5100$~K and $M\sim 0.56$ \msun. This is the expected
behavior discussed at the end of the previous subsection. As well, the
theoretical crystallization sequence (the two solid curves) sandwiches
remarkably well the observed distribution of stars to the right of the
convective coupling boundary in this lower panel. Hence, we find that
our C/O-core models provide a natural explanation for the observed clump
of low-mass, cool {\it Gaia} white dwarfs in the mass-effective
temperature diagram. These objects are likely currently undergoing a
significant slowdown in their cooling history due to the superposition
of latent heat and of extra thermal energy associated with the first
phase of convective coupling.

We note, in the present context, that there is a strong dependency
between the locations of isochrones and of the theoretical
crystallization sequence in a mass-effective temperature diagram and the
core composition of the models. This is illustrated by comparing the
lower with the upper panel of Figure \ref{crystallization_100pc_zoom},
where, in the latter case, isochrones and characteristic curves have
been plotted for models with pure C cores instead of C/O cores. Note
that these two sets of models are quite similar (same envelope
stratification, same physical assumptions), except for the core
composition. Given that carbon ions are less charged than their oxygen
counterparts, the former solidify at lower temperature in a dense
Coulomb plasma, given a density (equivalently, a total mass), it follows
that the theoretical crystallization sequence is shifted to lower
effective temperatures in the upper panel\footnote{Note, in this context, that
changing the core composition of evolutionary models of the kind has
little influence on the locations of the classical-to-quantum boundary
and none on the location of the convective coupling boundary.}. This
dependency on the core composition of models is most interesting and
directly implies that the core composition of {\it Gaia} white dwarfs
can be estimated, at least in bulk.  

The top panel of Figure \ref{crystallization_100pc_zoom} indicates that
the two solid curves, the crystallization sequence, do a significantly
poorer job of containing the high-density clump of stars than what is
depicted in the lower panel. This suggests that, as a whole, the {\it
  Gaia} white dwarfs are probably not made up of pure C. A similar and
complementary result would be obtained with pure O-core models, which
would predict, this time, a crystallization sequence too far to the left
of the observed clump. Of course, a proper statistical analysis, well
beyond the scope of this paper, is needed to infer correctly the most
probable bulk core composition for white dwarfs in the {\it Gaia}
sample. Nevertheless, we wish to attract the attention of the reader to
that very real possibility. Currently, we find that C/O-core models
perform better than pure C (and very likely pure O) models. The
statement that the core composition of {\it Gaia} white dwarfs, in bulk,
is more likely a mixture of C and O than pure C or pure O is hardly
a surprise. However, and to our knowledge, this is the first time that
such inference can be made on the basis of a very large number of stars.

\section{DISCUSSION}\label{sec:disc}

Despite its overall simplicity and straightforward approach, the
photometric technique is certainly not devoid of uncertainties, most
likely related to photometric calibration issues. For instance, the
question whether Pan-STARRS and SDSS magnitudes are precisely on the
AB magnitude system remains. As shown in our analysis, these two
photometric systems yield different effective temperatures, even when
we consider only the $griz$ filters common to both sets. To overcome
these calibration problems, \citet{HB2006} proposed a procedure where
empirical corrections are applied to various photometric systems,
which are calculated by comparing the observed magnitudes with those
predicted from spectroscopic determinations of $\Te$ and $\logg$ for
DA stars. However, this procedure {\it assumes} that the spectroscopic
solutions are accurate, which may not be the case according to the
analysis of GBB19. Nevertheless, our experiments
with various photometric systems indicate that the SDSS $ugriz$
photometric system, or a combination of SDSS $u$ with Pan-STARRS
$grizy$ photometry, probably yield the most accurate $\Te$
measurements. When the $u$ bandpass is omitted, however, the effective
temperatures are often underestimated (with the exception of SDSS
$griz$), and the corresponding stellar masses are underestimated as
well. But according to our results displayed in Figure
\ref{histo_mass_comp}, these mass differences remain small, $\sim$0.03
\msun\ on average, which is about the size of the mass uncertainty of
the spectroscopic method \citep{LBH05}.

The Hertzsprung-Russell diagram for white dwarfs presented in Figure
13 of GaiaHRD suggested that perhaps there was a problem with the
model atmospheres for DB white dwarfs, since the 0.6
\msun\ theoretical sequence failed to reproduce the observed sequence
in the $M_G$ vs $(G_{\rm BP}-G_{\rm RP})$ diagram. However, our
results displayed in Figure \ref{color_mag_mass} show that the
theoretical models for the DB stars ($\Te\gtrsim11,000$~K) overlap
perfectly with the observed sequence. The observed shift actually
occurs at lower temperatures where non-DA stars exist in the form of
DZ, DQ, or DC white dwarfs. By assuming pure helium compositions for
these objects, we found that the observed shift in absolute magnitude
corresponds to a shift in mass in a $M$ versus $\Te$ diagram (see
Figure \ref{correltm_MWDD_pureHe}). However, we also found that these
high masses could be significantly reduced to more normal masses
around 0.6 \msun\ if we include a small amount of hydrogen in the
atmosphere, the presence of which has been interpreted as the result
from the convective mixing of DA into non-DA white dwarfs at low
temperatures. Alternatively, the presence of other electron donors
such as carbon or metals would have the same effect. Consequently, the
population of massive white dwarfs interpreted as stellar mergers by
\citet{Kilic2018} is probably less important than previously
established.

\begin{figure}[t]
\centering 
\includegraphics[clip=true,trim=0.5cm 7cm 0cm 6cm,width=\linewidth]{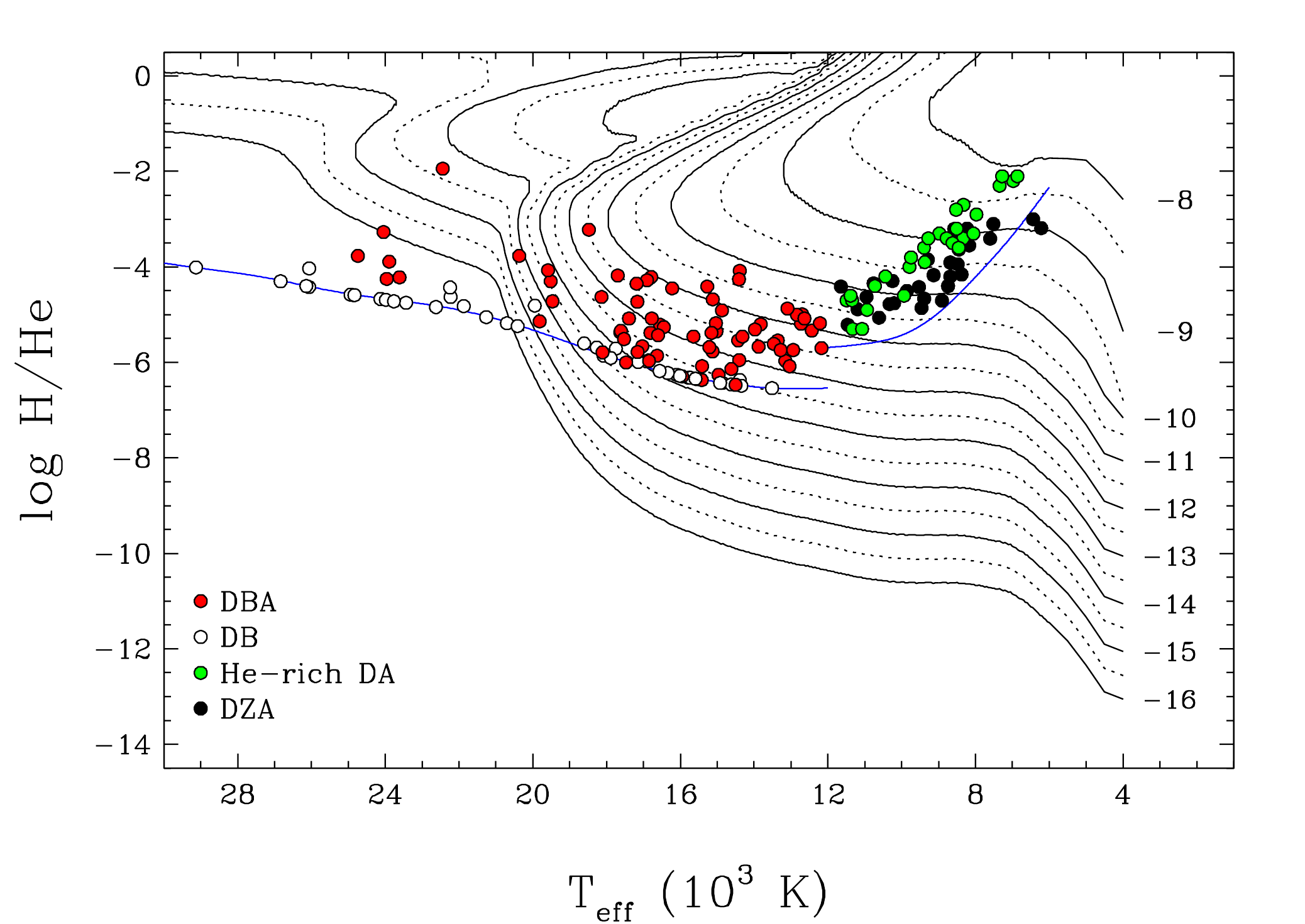}
\caption{Hydrogen-to-helium abundance ratio as a function of effective
  temperature for all DB, DBA, and cool, He-rich DA white dwarfs taken
  from the analysis of \citet{Rolland2018}; the DZA stars from
  \cite{dufour07} and \cite{Giammichele2012} are also displayed. The
  hydrogen detection limits at \ha\ are indicated by blue lines. Also
  reproduced are the predictions from our envelope calculations for
  homogeneously mixed models at 0.6 \msun\ for the ML2/$\alpha=0.6$
  version of the mixing-length theory. Each curve is labeled with the
  corresponding value of $\log M_{\rm H}/M_\odot$.
\label{seq_dilution}}
\end{figure}

According to our results displayed in Figure
\ref{correltm_MWDD_pureHe}, the number of cool ($11,000~{\rm
  K}>\Te>6000~{\rm K}$), pure helium-atmosphere white dwarfs with
normal masses must be extremely small. Now this is an odd result when
considering the chemical evolution of DB white dwarfs.  Figure 14 of
\citet{Rolland2018} --- reproduced here in Figure \ref{seq_dilution}
and extended to much lower effective temperatures --- shows the
hydrogen-to-helium abundance ratio as a function of effective
temperature for several DB, DBA, and cool, He-rich DA white dwarfs,
together with the predictions from envelope calculations for
homogeneously mixed models at 0.6 \msun.  By following a curve with a
constant value of total hydrogen mass (labeled as $\log M_{\rm
  H}/M_\odot$ in Figure \ref{seq_dilution}), one can predict the
evolution of the photospheric hydrogen abundance as a function of
time, or cooling temperature. The simulations indicate that a
significant fraction of DB white dwarfs should have hydrogen
abundances well below $\log\nh=-6$ once they reach temperatures below
$\Te=10,000$~K or so, and in particular for the ``pure'' DB stars with
no detectable traces of hydrogen (shown as white circles in Figure
\ref{seq_dilution}). Additional calculations, not discussed in our
paper, indicate that hydrogen abundances larger than $\log\nh\sim-6$
are required to affect the photometric masses above $\sim$8000
K. Since DB white dwarfs have normal masses around 0.6 \msun\ (see,
e.g., \citealt{Bergeron2011}), they should appear as $\sim$0.6 \msun,
pure helium white dwarfs in Figure \ref{correltm_MWDD_pureHe}. Yet,
there are none. We thus propose that another electron donor makes the
photometric masses of cooled off DB stars appear normal, namely
carbon. This implies that the progenitors of normal mass DQ stars are DB
white dwarfs. This proposition is perfectly in line with the well known
carbon dredge-up scenario which suggests a natural connection between
PG1159, DO, DB, and normal mass DQ white dwarfs (see, e.g.,
\citealt{Pelletier1986} ; \citealt{F2005}; \citealt{Dufour05}). 
\citet{B2007} last reviewed the question of the carbon pollution 
observed in the atmospheres of DQ stars and in some DB stars. They
proposed the existence of a convectively-driven cold wind in DB stars to
slow down the separation of C and O from He and, thus, maintain some
observable amounts of primordial C in their atmospheres. Alternatively,
the presence of radiative turbulence (of still unknown origin, however)
could be the source of competition against gravitational settling. 
Current thinking, based on the hydrodynamic simulations of
\citet{Tremblay2015}, rather suggests that overshooting at the base of the
superficial He convection zone in DO and DB white dwarfs could be the
actual competing mechanism. In any case, carbon does appear as a natural
electron donor in cooled off DB which produce normal mass DQ stars. 

One potential problem with the interpretation of the presence of trace
amounts of hydrogen in cool, helium-rich atmospheres is the behavior
at extremely low effective temperatures. Indeed, a typical hydrogen
abundance of $\log\nh=-5$ should produce an extremely strong infrared
flux deficiency at $\Te<4000$~K, resulting from collision-induced
absorptions by molecular hydrogen due to collisions with {\it helium},
as shown for instance in Figure 8 of \citet{Bergeron2002}. This
absorption is even more important than in pure hydrogen atmospheres,
and also manifests itself at much higher temperatures, and higher
luminosities.  Yet the color-magnitude diagram displayed in Figure
\ref{color_mag} shows no evidence for such a large population of cool
white dwarfs.  We turn again to Figure \ref{seq_dilution} for a
possible explanation of this lack of mixed H/He white dwarfs. The
simulations show that for most values of the total hydrogen mass, the
predicted hydrogen-to-helium abundance ratio remains nearly constant
below $\Te\sim11,000$~K, until the star reaches a temperature of
$\sim$6000~K, at which point the bottom of the mixed H/He convection
zone plunges deep into the star, resulting in photospheric hydrogen
abundances that are at least {\it two orders of magnitudes
  smaller}. Hence we suggest that most non-DA stars below 10,000~K
must have nearly pure helium atmospheres when they reach the end of
the cooling sequence. An alternative and more exotic explanation
proposed by \citet[][see their Section 6.3.2]{BRL97} is that some
non-DA stars could experience a sudden transition, where all the
hydrogen thoroughly diluted within the stellar envelope somehow makes
it back to the surface, thus transforming a mixed H/He white dwarf
into a hydrogen-dominated atmosphere white dwarf. This scenario would
be consistent with the fact that the coolest white dwarfs in the {\it
  Gaia} sample displayed in Figure \ref{correltm_PanSTARRS_100pc}
appear to have hydrogen-dominated atmospheres. More definitive
conclusions will have to wait until better photometric analyses of the
coolest white dwarfs become available, including photometric
measurements in the near infrared.

\acknowledgments PB would like to thank P.-E.~Tremblay for
enlightening discussions during his sabbatical stay at the University of
Warwick. GF acknowledges the contribution of the Canada Research Chair
Program. This work was supported in part by the NSERC Canada and by
the Fund FRQ-NT (Qu\'ebec). This work has made use of data from the
European Space Agency (ESA) mission {\it Gaia}
(\url{https://www.cosmos.esa.int/gaia}), processed by the {\it Gaia}
Data Processing and Analysis Consortium (DPAC,
\url{https://www.cosmos.esa.int/web/gaia/dpac/consortium}). Funding
for the DPAC has been provided by national institutions, in particular
the institutions participating in the {\it Gaia} Multilateral
Agreement. This research has made use of the NASA/ IPAC Infrared
Science Archive, which is operated by the Jet Propulsion Laboratory,
California Institute of Technology, under contract with the National
Aeronautics and Space Administration.

\clearpage

\bibliographystyle{aasjournal}
\bibliography{references}

\end{document}